\documentclass[twoside,twocolumn]{IEEEtran}

\newtheorem{theorem}{Theorem}
\newtheorem{lemma}{Lemma}
\newtheorem{corollary}{Corollary}

\newtheorem{remark}{Remark}
\newtheorem{definition}{Definition}
\newtheorem{claim}{Claim}
\newtheorem{example}{Example}

\newcommand{\ket}[1]{\mbox{$| #1 \rangle$}}
\newcommand{\bra}[1]{\mbox{$\langle #1 | $}}
\newcommand{\bracket}[2]{\mbox{$\langle #1 \mid #2 \rangle$}}
\newcommand{\neglog}[3]{\mbox{$\langle #1 \mid #2 \mid #3 \rangle$}}
\newcommand{\shar}[2]{\mbox{$ | #1 \rangle \langle #2 |$}}
\newcommand{\zero}{\ket{0}}
\newcommand{\one}{\ket{1}}

\newcommand{\norm}[1]{\mbox{$|| #1 ||$}}
{\protect\end{tabbing}}

\newcommand{\lea}{\stackrel{+}{<}}
\newcommand{\gea}{\stackrel{+}{>}}
\newcommand{\eqa}{\stackrel{+}{=}}

\begin{document}

\title{Quantum Kolmogorov Complexity Based on Classical Descriptions}
\author{Paul M.B. Vit\'anyi\thanks{Partially supported by the EU fifth framework
project QAIP, IST--1999--11234, the NoE QUIPROCONE IST--1999--29064,
the ESF QiT Programmme, and the EU Fourth Framework BRA
 NeuroCOLT II Working Group
EP 27150.
Part of this work was done
during the author's 1998 stay at Tokyo Institute of Technology,
Tokyo, Japan, as Gaikoku-Jin Kenkyuin at INCOCSAT, and
appeared in a preliminary version 
{\protect \cite{Vi00}}
%[19],
archived as quant-ph/9907035. Address:
CWI, Kruislaan 413, 1098 SJ Amsterdam, The Netherlands. Email:
{\tt paulv@cwi.nl}}
}

\markboth{IEEE Transactions on Information Theory}{Vit\'anyi: Quantum Kolmogorov Complexity Based on Classical Descriptions}

\maketitle

\begin{abstract}
We develop a theory of the algorithmic information in bits contained
in an individual pure quantum state.
This extends classical Kolmogorov complexity to the quantum domain
retaining classical descriptions.
Quantum Kolmogorov complexity
coincides with the classical Kolmogorov complexity 
on the classical domain.
Quantum Kolmogorov complexity is upper
bounded and can be effectively approximated from above under certain
conditions.
With high probability a quantum object is incompressible. 
Upper- and lower bounds of the quantum complexity of multiple
copies of individual pure quantum states are derived
and may shed some light on
the no-cloning properties of quantum states.
In the quantum situation complexity 
is not sub-additive.
We discuss some relations with ``no-cloning''
and ``approximate cloning'' properties.
\end{abstract}

\begin{keywords}
Algorithmic information theory, quantum;
classical descriptions of quantum states;
information theory, quantum;
Kolmogorov complexity, quantum; 
quantum cloning.
\end{keywords}

\section{Introduction}
\PARstart{Q}{uantum} information theory, the quantum mechanical analogue
of classical information theory \cite{CT91}, is experiencing
a renaissance \cite{BS98} due to the rising interest in
the notion of quantum computation and the possibility
of realizing a quantum computer \cite{NC00}.
While Kolmogorov complexity \cite{Ko65} is the accepted absolute measure
of information content in a {\em individual classical} finite object,
a similar absolute notion is needed for the information
content of an {\em individual} pure {\em quantum} state.
One motivation is to extend probabilistic quantum information theory
to Kolmogorov's absolute individual notion. Another reason is
to try and duplicate the success of classical Kolmogorov complexity 
as a general proof method in applications ranging from combinatorics
to the analysis of algorithms, and from pattern recognition to
learning theory \cite{LiVi97}. We propose a theory of quantum Kolmogorov 
complexity based on classical descriptions and derive the results
given in the abstract. A preliminary partial version appeared as \cite{Vi00}.

What are the problems and choices to be made 
developing a theory of quantum Kolmogorov complexity?
Quantum theory assumes that every complex vector of unit length
represents a realizable pure quantum state \cite{Pe95}. 
%Two complex vectors pointing in
%the same direction give the same quantum state and consequently
%the theory considers only vectors of unit length.
There arises the question
of how to design the equipment that prepares such a pure
state. While there are continuously many pure states in 
a finite-dimensional complex vector space---corresponding to all vectors of
unit length---we can finitely
describe only a countable subset. Imposing effectiveness on
such descriptions leads to constructive procedures.
The most general such procedures satisfying universally agreed-upon
logical principles of effectiveness are quantum Turing machines, \cite{BV97}.
To define quantum Kolmogorov complexity 
by way of quantum Turing machines leaves
essentially two options:
\begin{enumerate}
\item
We want to describe every quantum superposition exactly; or
\item
we want to take
into account the number of bits/qubits in the specification
as well the accuracy of the quantum state produced. 
\end{enumerate}
We have to deal with three problems:
\begin{itemize}
\item
There are continuously many quantum Turing machines;
\item
There are continuously many pure quantum states;
\item
There are continuously many qubit descriptions.
\end{itemize}
There are uncountably many quantum Turing machines
only if we allow arbitrary real rotations in the definition of 
machines. Then, a quantum Turing machine can only be universal
in the sense that it can approximate the computation of an
arbitrary machine, \cite{BV97}. In descriptions using universal
quantum Turing machines
we would have to account for the closeness of approximation,
the number of steps required to get this precision, and the like.
In contrast, if we fix the rotation
of all contemplated machines to a single primitive rotation $\theta$ 
with $\cos \theta = \frac{3}{5}$ and $\sin \theta = \frac{4}{5}$, then there
are only countably many Turing machines and the universal machine
simulates the others exactly \cite{ADH97}. 
Every quantum
Turing machine computation, using arbitrary real rotations
to obtain a target pure quantum state,
can be approximated to every precision by machines with fixed
rotation $\theta$ but in general cannot be simulated 
exactly---just like in the case of the simulation of
arbitrary quantum Turing machines by a universal
quantum Turing machine.  Since exact simulation is impossible
by a fixed universal quantum Turing machine anyhow, but arbitrarily
close approximations are possible by Turing machines using
a fixed rotation like $\theta$, we are motivated to fix 
$Q_1 , Q_2 , \ldots$ as a standard enumeration of
quantum Turing machines using only rotation $\theta$.

Our next question is whether we want programs 
(descriptions) to be in classical bits
or in qubits?
The intuitive notion of computability requires
the programs to be classical. Namely, to prepare a quantum state
requires a physical apparatus that ``computes'' this quantum state
from classical specifications. 
Since such specifications have effective descriptions,
every quantum state that can be prepared can 
be described effectively in descriptions consisting of classical bits.
Descriptions consisting of arbitrary pure quantum states
allows noncomputable (or hard to compute)
information to be hidden in the bits of
the amplitudes.
In Definition~\ref{def.pqscomp} we call a pure quantum state {\em directly
computable} if there is a (classical) program such that
the universal quantum Turing machine computes that state from
the program and then halts in an appropriate fashion.
In a computational setting we naturally
require that directly computable pure quantum states can be 
prepared.
By repeating the preparation we can obtain
arbitrarily many copies of the pure quantum state.

If descriptions are not effective then we are not going to use them in our
algorithms except possibly on inputs from an ``unprepared''
origin. Every quantum state used in a quantum computation
arises from some classically preparation or is possibly
captured from some unknown origin. If the latter, then we can consume
it as conditional side-information or an oracle. 

Restricting ourselves to an effective enumeration of
quantum Turing machines and classical descriptions
to describe by approximation continuously many pure quantum states is
reminiscent of the construction of continuously many real  numbers
from Cauchy sequences of rational numbers, the rationals being 
effectively enumerable. 

{\bf Kolmogorov complexity:}
We summarize some basic definitions in Appendix~\ref{app.A}
(see also this journal \cite{ViLi00})
in order to establish notations and recall the notion
of shortest effective descriptions. More details can be 
found in the textbook \cite{LiVi97}.
Shortest effective descriptions are ``effective''
in the sense that they are programs:
 we can compute the described objects from them.
Unfortunately, \cite{Ko65}, there is no algorithm that computes
the shortest program and then halts, that is,
there is no general method to compute the length of a shortest description
(the Kolmogorov complexity) from the object
being described. This obviously impedes actual use. Instead, one
needs to consider computable approximations to shortest descriptions,
for example
by restricting the allowable approximation time. Apart
from computability and approximability, there is another
property of descriptions that is important to us. A set of
descriptions is {\em prefix-free} if no description is a 
proper prefix of another description. Such a set is called
a {\em prefix code}. Since a code message consists of concatenated
code words, we have to parse it into its constituent code words
to retrieve the encoded source message. If the code is 
{\em uniquely decodable}, then every code message can be decoded in only
one way. The importance of prefix-codes  stems from the fact that
(i) they are uniquely decodable from left to right without backing up,
and (ii) for every uniquely decodable code there is a prefix code with
the same length code words. Therefore, we can restrict ourselves
to prefix codes.
In our setting we require 
the set of programs to be prefix-free and hence to be a prefix-code
for the objects being described. It is well-known that with every
prefix-code there corresponds a probability distribution $P(\cdot)$
such that the prefix-code is a
Shannon-Fano code
\footnote{In what follows, ``$\log$'' denotes the binary logarithm.
%% ``$\lfloor r \rfloor$'' is the greatest integer $q$
%% such that $q \leq r$.
}
that assigns prefix code length $l_x = - \log P(x)$ to $x$---irrespective
of the regularities in $x$. For example, with the uniform
distribution $P(x)=2^{-n}$ on the set of $n$-bit source words,
the Shannon-Fano code word length of an
all-zero source word equals the code word length of a truly
irregular source word. The Shannon-Fano code gives an expected
code word length close to the entropy, and, by Shannon's Noiseless
Coding Theorem, it possesses the optimal expected code word length.
But the Shannon-Fano code is not optimal for individual
elements: it does not take advantage of the regularity in some
elements to encode those shorter.
In contrast, one can view the Kolmogorov complexity $K(x)$
as the code word length of the shortest program $x^*$ for $x$,
the set of shortest programs consitituting the Shannon-Fano code of
the so-called ``universal distribution'' ${\bf m}(x) = 2^{-K(x)}$.
The code consisting of the shortest 
programs has the remarkable property
that it achieves (i) an expected code length that is about
optimal since it is close to the entropy,
{\em and simultaneously}, (ii) every individual object is coded as short as is
effectively possible, that is, squeezing out all regularity.
In this sense the set of shortest programs constitutes the optimal
effective Shannon-Fano code, induced by the optimal effective distribution
(the universal distribution).

{\bf Quantum Computing:}
We summarize some basic definitions in Appendix~\ref{app.B}
in order to establish notations and briefly review the notion
of a quantum Turing machine computation. See also this journal's survey
\cite{BS98} on quantum information theory. More details
can be found in the textbook \cite{NC00}. Loosely speaking,
like randomized computation is a generalization of deterministic
computation, so is quantum computation a generalization of
randomized computation. Realizing a mathematical random source to
drive a random computation is, in its ideal form, presumably
impossible (or impossible to certify) in practice. Thus, in
applications an algorithmic random number generator is used. 
Strictly speaking this invalidates the analysis 
based on mathematical randomized 
computation. As John von Neumann \cite{vN51}
put it: ``Any one who considers 
arithmetical methods of producing random digits is,
of course, in a state of sin. For, as has been pointed out several
times, there is no such thing as a random number---there are
only methods to produce random numbers, and a strict arithmetical
procedure is of course not such a method.'' In practice
randomized computations reasonably satisfy theoretical analysis. 
In the quantum computation setting, the practical
problem is that the ideal coherent superposition
cannot really be maintained during computation but
deteriorates---it decoheres. In our analysis we abstract from
that problem and one hopes that in practice anti-decoherence techniques
will suffice to approximate the idealized performance sufficiently.

We view a quantum Turing machine as a generalization
of the classic probabilistic (that is, randomized) Turing
machine. The probabilistic Turing machine computation
follows multiple computation paths in parallel,
each path with a certain
associated probability. The quantum Turing machine computation
follows multiple computation paths in parallel, but now every path
has an associated complex probability amplitude.
If it is possible to reach the same state via different paths,
then in the probabilistic case the probability of observing
that state is simply the sum of the path probabilities. 
In the quantum case it is the squared norm of the summed 
path probability amplitudes. Since the probability amplitudes
can be of opposite sign, the observation probability can vanish; 
if the path probability
amplitudes are of equal sign then the observation probability
can get boosted since it is the {\em square}
of the sum norm. While this generalizes the probabilistic aspect,
and boosts the computation power through the phenomenon
of interference between parallel computation paths,
there are extra restrictions vis-a-vis probabilistic
computation in that the quantum evolution
must be unitary.

{\bf Quantum Kolmogorov Complexity:}
We define the Kolmogorov complexity of a pure quantum state
as the length of the shortest two-part code consisting
of a classical program to compute an approximate pure quantum state and
the negative log-fidelity of the approximation to the target quantum state.
We show that the resulting quantum Kolmogorov complexity
coincides with the classical self-delimiting complexity on the domain
of classical objects; and that certain properties that we love and cherish
in the classical Kolmogorov complexity are shared by the new quantum
Kolmogorov complexity: quantum Kolmogorov complexity of an $n$-qubit
object is upper bounded by about $2n$; it
is not computable
but can under certain conditions
 be approximated from above by a computable process;
and with high probability a quantum object is incompressible.
We may call this quantum Kolmogorov complexity the 
{\em bit complexity} of a pure
quantum state $\ket{\phi}$ (using Dirac's ``ket'' notation) and denote it 
by $K( \ket{\phi})$.  
From now on, we will denote by $\lea$ an inequality to within an
additive constant, and by $\eqa$ the situation when both $\lea$ and
$\gea$ hold. For example,
we will show that, for $n$-qubit states $\ket{\phi}$,
the complexity satisfies $K( \ket{\phi} \mid n ) \lea 2n$.
For  certain restricted pure quantum states, quantum kolmogorov
complexity satisfies the sub-additive property: 
$K(\ket{\phi , \psi}) \lea K(\ket{\phi})+ K(\ket{\psi} \mid \ket{\phi})$.
But, in general, quantum Kolmogorov complexity
is {\em not} sub-additive.
Although ``cloning'' of non-orthogonal states 
is forbidden in the quantum setting \cite{WZ82,Di82}, 
$m$ copies of the same quantum state have combined complexity
that can be considerable lower
than $m$ times the complexity of a single copy.
In fact, quantum Kolmogorov complexity appears to enable us to 
express
and partially quantify
``non-clonability'' and
``approximate clonability'' of {\em individual} pure quantum states.

{\bf Related Work:} In the classical situation there are several variants
of Kolmogorov complexity that are very meaningful in their
respective settings: plain Kolmogorov complexity, prefix complexity,
monotone complexity, uniform complexity,
negative logarithm of universal measure, and so on \cite{LiVi97}.
It is therefore not surprising that in the more complicated situation
of quantum information several different choices of complexity 
can be meaningful and unavoidable in different settings. 
Following the preliminary version \cite{Vi00} of this work there have been
alternative proposals:

{\bf Qubit Descriptions:}
The most straightforward
way to define a notion of quantum Kolmogorov complexity
 is to consider the shortest effective qubit description
of a pure quantum state which is studied in \cite{BDL00}. 
(This {\em qubit complexity} can also be 
formulated in terms of the conditional version
of bit complexity as in \cite{Vi00}.)
An advantage of qubit complexity is that the
upper bound on the complexity of a pure
quantum state is immediately given by the number of qubits involved in the
literal description of that pure quantum state. Let us denote the
resulting qubit complexity of a pure quantum state $\ket{\phi}$  by
$KQ(\ket{\phi})$.

While it is clear that (just as with the previous aproach)
the qubit complexity is not computable, it is unlikely
that one can approximate the qubit complexity from above by
a computable process in some meaningful sense. 
In particular, the dovetailing approach
we used in our approach now doesn't seem applicable due
to the non-countability of the potentential qubit program candidates.
The quantitative incompressibility properties are much like the
classical case (this is important for future applications).
There are some interesting exceptions in case of objects consisting
of multiple copies related to the ``no-cloning'' property of quantum 
objects, \cite{WZ82,Di82}. Qubit complexity does not
satisfy the sub-additive property, 
and a certain version of it (bounded fidelity)
is bounded above by the von Neumann entropy.

{\bf Density Matrices:}
In classical algorithmic information theory it turns out that
the negative logarithm of the ``largest'' 
probability distribution
effectively approximable from below---the universal distribution---coincides
with the self-delimiting Kolmogorov complexity. In \cite{Ga00} G\'acs
defines two notions of complexities based on the negative logarithm
of the ``largest'' density matrix ${\bf \mu}$ effectively
approximable from below. 
There arise two different complexities of $\ket{\phi}$
based on whether we take the logarithm inside 
as $KG (\ket{\phi})= - \neglog{ \phi }{ \log {\bf \mu}}{ \phi }$ or outside as
$Kg( \ket{\phi}) = - \log \neglog{ \phi }{ {\bf \mu} }{ \phi }$.
It turns out
 that $Kg(\ket{\phi}) \lea KG( \ket{\phi})$. This approach
serves to compare the two approaches above: It was
shown that $Kg (\ket{\phi})$ is within a factor four of $K(\ket{\phi})$;
that $KG(\ket{\phi})$ essentially is a lower bound on $KQ(\ket{\phi})$
and an oracle version of $KG$ is essentially an upper bound on qubit
complexity $KQ$. Since qubit complexity is trivially $\lea n$
and it was shown that bit complexity is typically close to $2n$,
at first glance this leaves the possibility that the two
complexities are within a factor two of each other. 
This turns out to be not the case since it was shown that the $Kg$ complexity
can for some arguments be much smaller than the $KG$ complexity,
so that the bit complexity is in these cases also much smaller than 
the qubit complexity. As \cite{Ga00} states: this is due to the permissive
way the bit complexity deals with approximation.
The von Neumann entropy of a computable density matrix
is within an additive constant (the complexity of the program
computing the density matrix) of a notion of average complexity.
The drawback of density matrix based complexity is that we seem to
have lost the direct relation with a meaningful interpretation
in terms of description length: a crucial
aspect of classical Kolmogorov complexity in most applications \cite{LiVi97}.

{\bf Real Descriptions:}
A version of quantum Kolmogorov complexity briefly
considered in \cite{Vi00} uses
computable real parameters to describe the pure quantum state
with complex probability amplitudes.
This requires two reals per complex probability amplitude, that is,
for $n$ qubits one requires $2^{n+1}$ real numbers in the worst case.
A real number is computable if there is a fixed program that
outputs consecutive bits of the binary expansion of the number forever.
Since every computable real number may require a separate program,
a computable $n$-qubit pure state may require $2^{n+1}$ finite programs.
Most $n$-qubit pure states have parameters that 
are noncomputable and increased 
precision will require increasingly long programs. For example,
if the parameters are recursively enumerable (the positions of the ``1''s
in the binary
expansion is a recursively enumerable set),
then a $\log k$ length program per parameter,
to achieve $k$ bits precision per recursively enumerable real, is sufficient
and for some recursively enumerable reals also necessary.
In certain contexts where the approximation of the real parameters
is a central concern, 
such considerations may be useful. 
While this approach does not allow the development of a clean
theory in the sense of the previous approaches, it can be directly
developed in terms of algorithmic thermodynamics---an extension
of Kolmogorov complexity to randomness of infinite sequences
(such as binary expansions of real numbers)
in terms of coarse-graining and sequential Martin-L\"of tests, 
analogous to the classical case in \cite{Ga94,LiVi97}.
But this is outside the scope of the present paper.

\section{Quantum Turing Machine Model}
\label{sect.model}
We assume the notation and definitions in Appendices~\ref{app.A}, \ref{app.B}.
Our model of computation is a quantum
Turing machine equipped with a input tape 
that is one-way infinite
with the classical input (the program)
in binary left adjusted from the beginning. 
We require that the input tape is read-only from left-to-right 
without backing up. This automatically yields a
property we require in the sequel: 
The set of halting programs is prefix-free.
Additionaly, the machine
contains a one-way infinite
work tape containing qubits, a one-way infinite auxiliary
tape containing qubits, and a one-way infinite output tape
containing qubits. Initially, the input tape contains
a classical binary program $p$, and all (qu)bits
of the work tape, auxiliary tape, and output
tape qubits are set to $\ket{0}$. In case the Turing machine
has an auxiliary input (classical or quantum) then initially
the leftmost qubits of the auxiliary tape contain this input. 
A quantum Turing machine $Q$ with classical program $p$ and 
auxiliary input $y$ computes until it halts with output
$Q(p, y)$ on its output tape or it computes forever.
Halting is a more complicated matter here than in the classical case
since quantum Turing machines are reversible, which means that
there must be an ongoing evolution with nonrepeating configurations.
There are various ways to resolve this problem \cite{BV97} and we
do not discuss this matter further. We only consider quantum 
Turing machine that do not modify the output tape 
after halting.  Another---related---problem is that
after halting the quantum state on the output tape may be ``entangled''
with the quantum state of the remainder of the machine, that is,
the input tape, the finite control, the work tape, and the auxilliary
tape. This has the effect that
the output state viewed in isolation may not be a pure
quantum state but a mixture of pure quantum states.
This problem does not arise if the output and the remainder of the machine
form a tensor product 
so that the output is un-entangled with the
remainder. The results in this paper are invariant
under these different assumptions, but considering
output entangled with the remainder of the machine
complicates formulas and calculations.
Correspondingly, we restrict consideration to 
outputs that form a tensor product with the remainder
of the machine, with the understanding that the same results hold
with about the same proofs if we choose the other option---except
in the case of Theorem~\ref{theo.appr} item (ii), see the pertinent
caveat there. 
Note that the Kolmogorov complexity based on 
entangled output tapes is at most (and conceivably less than)
the Kolmogorov complexity
based on un-entangled output tapes.
\begin{definition}
Define the {\em output} $Q(p, y)$ of a
quantum Turing machine $Q$ with classical program $p$ and
auxiliary input $y$ as the pure quantum state $\ket{\psi}$
resulting of $Q$ computing until it halts with output
$\ket{\psi}$ on its ouput tape. Moreover, $\ket{\psi}$
doesn't change after halting, and it
is un-entangled with the remainder of $Q$'s configuration.
We write $Q(p,y) < \infty$.
If there is no such $\ket{\psi}$ then $Q(p, y)$ is undefined and
we write $Q(p,y) = \infty$. 
By definition the input tape is read-only from left-to-right
without backing up: therefore the set of {\em halting
programs} ${\cal P}_y  = \{p: Q(p , y) < \infty \}$ is {\em prefix-free}:
no program in ${\cal P}_y$ is a proper prefix of another program in
${\cal P}_y$. Put differently, the Turing machine scans all of a
halting program $p$ but never scans the bit following the last
bit of $p$: it is {\em self-delimiting}.
\end{definition} 

We fix the rotation
of all contemplated machines to a single primitive rotation $\theta$
with $\cos \theta = \frac{3}{5}$ and $\sin \theta = \frac{4}{5}$. There
are only countably many such Turing machines. 
Using a standard ordering,
we fix $Q_1 , Q_2 , \ldots$ as a standard enumeration of
quantum Turing machines using only rotation $\theta$.
By \cite{ADH97}, there is a universal machine $U$ in this enumeration
that simulates the others exactly: $U(1^i0p,y) = Q_i (p,y)$,
for
all $i,p,y$. (Instead of the many-bit encoding $1^i0$ for $i$
we can use a shorter self-delimiting code like $i'$ in Appendix~\ref{app.A}.)
As noted in the Introduction, every quantum
Turing machine computation using arbitrary real rotations
can be approximated to arbitrary precision by machines with fixed
rotation $\theta$ but in general cannot be simulated
exactly.

%\begin{remark}
%\rm
%This contrasts with the approach, originally taken
%by Kolmogorov \cite{Ko65}, were the input $p$ is delimited
%by distinguished markers. Then the Turing machine always knows where
%the input ends. In the self-delimiting case the endmarker must be
%implicit in the halting program $p$ itself. This encoding of the
%endmarker carries an inherent penalty in the form of increased length:
%typically a prefix code of an $n$-length binary string has length
%about $n+ \log n + 2 \log \log n$ bits, \cite{LiVi97}.
%\end{remark}
\begin{remark}
\rm
There are two possible interpretations for the computation relation
$Q(p, y) = \ket{x}$. In the narrow interpretation
we require that $Q$ with $p$ on the input tape
and $y$ on the conditional tape halts with $\ket{x}$
on the output tape. In the wide interpretation we can
define pure quantum states by requiring
that for every precision parameter $k > 0$ the computation
of $Q$ with $p$ on the input tape
and $y$ on the conditional tape,
with $k$ on a special  new tape where the precision is to be supplied,
halts with $\ket{x'}$
on the output tape and $\norm{\bracket{x}{x'}}^2 \geq 1-1/2^k$.
Such a notion of ``computable''
or ``recursive'' pure quantum states is similar to Turing's
notion of ``computable numbers.'' 
In the remainder of this section we use the narrow interpretation.
\end{remark}

\begin{remark}
\rm
As remarked in \cite{Ga00}, the notion of a quantum computer is not essential
to the theory here or in \cite{BDL00,Ga00}.
Since the computation time of the machine is not limited
in the theory of description complexity as developed here,
a quantum computer can be simulated by a classical computer to every
desired degree of precision. 
We can rephrase everything in terms
of the standard enumeration of 
$T_1 , T_2 , \ldots$ of classical Turing machines. Let $\ket{x}
= \sum_{i=0}^{N-1} \alpha_i \ket{e_i}$ ($N=2^n$)
be an $n$-qubit state.
We can write $T(p) = \ket{x}$
if $T$ either outputs

(i) algebraic definitions of the coefficients of $\ket{x}$ (in case these
are algebraic), or

(ii) a sequence of approximations 
$(\alpha_{0,k} , \ldots , \alpha_{N-1, k})$ for $k=1,2, \ldots$
where $\alpha_{i,k}$ is an algebraic approximation of $\alpha_i$
to within $2^{-k}$.
\end{remark}

\section{Classical Descriptions of Pure Quantum States}

The complex
quantity $\bracket{x}{z}$ is the inner product of vectors $\bra{x}$ 
and $\ket{z}$.
Since pure quantum states $\ket{x}, \ket{z}$ have unit length, 
$\norm{\bracket{x}{z}} =  | \cos \theta |$  where $\theta$ is the
angle between vectors $\ket{x}$ and $\ket{z}$.
The quantity $\norm{\bracket{x}{z}}^2$, the {\em fidelity}
between $\ket{x}$ and $\ket{z}$, is a measure of how ``close''
or ``confusable'' the vectors $\ket{x}$ and $\ket{z}$
are. It is the probability of outcome
$\ket{x}$ being
measured from state $\ket{z}$. 
Essentially,  we project $\ket{z}$ on outcome
$\ket{x}$ using projection $\shar{x}{x}$ resulting in
$\bracket{x}{z} \ket{x}$.

\begin{definition}\label{def.qkc}
\rm
The {\em (self-delimiting) complexity} of $\ket{x}$ 
with respect to quantum Turing machine $Q$
with $y$ as conditional input given for free is 
\begin{equation}\label{eq.qkc}
K_Q (\ket{x} \mid y ) = 
\min_{p} \{ l(p) + \lceil - \log \norm{\bracket{z}{x}}^2 \rceil : 
Q(p, y) = \ket{z} \}
\end{equation}
where $l(p)$ is the number of bits in the program $p$,
auxiliary $y$ is an input (possibly quantum) state, and 
$\ket{x}$ is the target state that one is
trying to describe. 
\end{definition}

Note that $\ket{z}$ is
the quantum state produced by the computation $Q(p, y)$, 
and therefore, given $Q$ and $y$, completely determined by $p$.
Therefore, we obtain the minimum of the right-hand side of
the equality by minimizing over $p$ only.
We call the $\ket{z}$ that minimizes the right-hand side the
{\em directly computed part} of $\ket{x}$ while 
$ \lceil - \log \norm{\bracket{z}{x}}^2 \rceil$ is the 
{\em approximation part}.

Quantum Kolmogorov complexity is the sum
of two terms: the first term is the integral length
of a binary program, and the second term,
the minlog probability term, corresponds to the length
of the corresponding code word in the Shannon-Fano code associated
with that probability distribution, see for example \cite{CT91},
and is thus also
expressed in an integral number of bits.
Let us consider this relation more closely:
For a quantum system 
$\ket{z}$
the quantity $P(x)= \norm{\bracket{z}{x}}^2$ is the probability that
the system passes a test for $\ket{x}$, and vice versa.
The term $\lceil - \log \norm{\bracket{z}{x}}^2 \rceil$ can be viewed
as the  
code word length to redescribe $\ket{x}$, given $\ket{z}$ 
and an orthonormal basis with $\ket{x}$ as one of the basis vectors,
using
the Shannon-Fano prefix code.
This works as follows: Write $N=2^n$. For every state $\ket{z}$ in 
$(2^n)$-dimensional Hilbert space
with basis vectors ${\cal B} = \{ \ket{e_0}, \ldots , \ket{e_{N-1}}\}$ we have
$\sum_{i=0}^{N-1} \norm{ \bracket{e_i }{z}}^2 =1$. If the basis has
$\ket{x}$ as one of the basis vectors, then we can
consider $\ket{z}$ as a random variable that assumes value $\ket{x}$
with probability $\norm{\bracket{x}{z}}^2$. The Shannon-Fano code word
for $\ket{x}$ in the probabilistic ensemble
$\left({\cal B} , (\norm{\bracket{e_i}{z}}^2)_i \right)$ is 
based on the probability $\norm{\bracket{x}{z}}^2$ of $\ket{x}$,
given $\ket{z}$, and has length  
$\lceil - \log \norm{\bracket{x}{z}}^2 \rceil$. Considering a canonical
method of constructing an orthonormal basis 
${\cal B} = \ket{e_0}, \ldots, \ket{e_{N-1}}$ 
from a given basis
vector, we can choose ${\cal B}$ such that 
$K({\cal B}) \eqa \min_i \{ K(\ket{e_i}) \}$.
The Shannon-Fano code is appropriate for our purpose since it is optimal
in that it achieves the least expected code word
length---the expectation taken over the probability of the
source words---up to 1 bit by Shannon's Noiseless Coding Theorem.
As in the classical case the quantum Kolmogorov
complexity is an integral number.

The main property required to be able to develop a meaningful
theory is that our definition satisfies
a so-called {\em Invariance Theorem} (see also Appendix
\ref{app.A}).
Below we use ``$U$'' to denote a special type
of universal (quantum) Turing machine
rather than a unitary matrix.
\begin{theorem}[Invariance]\label{theo.inv}
There is a universal machine 
$U$, such that for all machines $Q$,
there is a constant $c_Q$ (the length of the
description of the index of $Q$ in the enumeration),
such that for all quantum states $\ket{x}$ and all 
auxiliary inputs $y$ we have:
\[
K_U (\ket{x} \mid y) \leq K_Q (\ket{x}\mid y) + c_Q. 
\]
\end{theorem}

\begin{proof}
Assume that the program $p$ that minimizes the right-hand side
of (\ref{eq.qkc}) is $p_0$ and the computed $\ket{z}$ is $\ket{z_0}$:
\[
K_Q (\ket{x} \mid y ) =
l(p_0) + \lceil - \log \norm{\bracket{z_0}{x}}^2 \rceil .
\]
There is a universal quantum Turing machine $U$ in the standard enumeration
$Q_1 , Q_2, \ldots$ such that for every quantum Turing machine
$Q$ in the enumeration there is a self-delimiting program $i_Q$
(the index of $Q$) and $U(i_Q p , y) = Q(p,y)$ for all $p,y$:
if $Q(p,y) = \ket{z}$ then $U(i_Q p , y)= \ket{z}$. In particular,
this holds for $p_0$ such that $Q$ with auxiliary
input $y$ halts with output $\ket{z_0}$. 
But $U$ with auxiliary
input $y$ halts on input $i_Q p_0$ also with output $\ket{z_0}$.
Consequently, the program $q$ that minimizes the
right-hand side of (\ref{eq.qkc}) with $U$ substituted for $Q$,
and computes $U(q,y) = \ket{u}$ for some state $\ket{u}$ possibly
different from $\ket{z}$, satisfies
\begin{eqnarray*}
K_U( \ket{x} \mid y) & = & l(q) + \lceil - \log \norm{\bracket{u}{x}}^2 \rceil \\
& \leq &
l(i_Q p_0) + \lceil - \log \norm{\bracket{z_0}{x}}^2 \rceil .
\end{eqnarray*}
Combining the two displayed inequalities, and
setting $c_Q = l(i_Q)$, proves the theorem.
\end{proof}

The key point is not that the universal Turing machine
viewed as description method
does necessarily give the shortest description in each
case, but that no other effective description method can improve
on it infinitely often by more than a fixed constant.
For %
\it every %
\rm pair $U ,  U ' $
of universal Turing machines as in the proof of
Theorem~\ref{theo.inv}, there
is a fixed constant $c_{{U} , U' }$,
depending only on $U$ and $U'$, such
that for all $\ket{x}, y$ we have:
\[
|K_{U} (\ket{x} \mid  y) - K_{U'} (\ket{x}\mid y)|   \leq   c_{U , U ' } .
\]
To see this, substitute $U'$ for $Q$ in (\ref{eq.qkc}),
and, conversely, substitute $U'$ for $U$ and $U$ for $Q$ in (\ref{eq.qkc}),
and combine the two resulting
inequalities.
While the complexities according to $U$ and $U'$
are not exactly equal, they are %
\it equal up to a fixed constant %
\rm for all $\ket{x}$ and $y$.
Therefore, one or the other fixed choice of reference universal machine $U$
yields resulting complexities that are in a fixed constant
enveloppe from each other for all arguments.

Programmers are generally aware that programs for
symbolic manipulation tend to be shorter when they are expressed
in the LISP programming language than if they are expressed
in FORTRAN, while for numerical calculations the opposite
is the case. Or is it? The Invariance Theorem in fact shows
that to express an algorithm succinctly in a program,
it does not matter which programming language we use---
up to a fixed additive constant (representing the length
of compiling programs from either language into the other
language) that
depends only on the two programming languages compared.
For further discussion of effective optimality and invariance
see \cite{LiVi97}.

\begin{definition}
\rm
We fix once and for all a 
{\em reference universal quantum Turing machine} $U$
and define the {\em quantum Kolmogorov complexity} as
\begin{eqnarray*}
&& K (\ket{x} \mid  y) = K_U (\ket{x}\mid y), \\
&& K (\ket{x}) = K_U (\ket{x} \mid  \epsilon ), 
\end{eqnarray*}
where $\epsilon$ denotes the absence of conditional
information.
\end{definition}

The definition is continuous:
If two quantum states are very close then their quantum Kolmogorov
complexities are very close. Furthermore, since we can approximate
every (pure quantum) state $\ket{x}$ to arbitrary closeness, \cite{BV97},
in particular, for every constant $\epsilon > 0$
we can compute a (pure quantum) state $\ket{z}$
such that 
 $\norm{\bracket{z}{x}}^2 > 1-\epsilon$.
One can view this as the probability of obtaining the possibly 
noncomputable outcome $\ket{x}$
when executing projection $\shar{x}{x}$ on $\ket{z}$ 
and measuring outcome $\ket{x}$.
For this definition to be
useful it should satisfy:
\begin{itemize}
\item
The complexity of a pure state that can be directly computed should be the
length of the shortest program that computes that state. (If the
complexity is less then this may lead to discontinuities when we restrict
quantum Kolmogorov complexity to the domain of classical objects.)
\item
The quantum Kolmogorov complexity of a classical object should
equal the classical Kolmogorov complexity of that object (up to
a constant additive term).
\item
The quantum Kolmogorov complexity of a quantum object should
have an upper bound. (This is necessary for the complexity
to be approximable from above, even if the quantum object is
available in as many copies as we require.)
\item
Most objects should be ``incompressible'' in terms of quantum
Kolmogorov complexity.
\item
In the classical case the average self-delimiting
Kolmogorov complexity of the discrete set of all $n$-bit strings
under some distribution
equals the Shannon entropy up to an additive constant depending
on the complexity of the distribution concerned. In our
setting we would like to know
the relation between the expected $n$-qubit quantum Kolmogorov
complexity, the expectation taken over a computable
(semi-)measure over the continuously many $n$-qubit states,
with von Neumann entropy. Perhaps the continuous set can be restricted
to a representative discrete set. We have no results along these lines.
One problem may be that in the quantum
situation there can be many different mixtures
of pure quantum states that give rise to the same density matrix,
and thus have the same von Neumann entropy. It is possible
that the average Kolmogorov complexities of different
mixtures with the same density matrix (or density matrices with
the same eigenvalues) are also different (and therefore
not all of them can be equal to the single fixed 
von Neumann entropy which depends
only on the eigenvalues). 
%A particular degenerate
%case where they are equal is if we consider the average Kolmogorov complexity
%of the set of states in an orthonormal basis. Since it is known that
%the eigen values of the density matrix of a mixture are invariant under 
%a change of orthonormal basis, in particular the von Neumann entropy
%doesn't change. But this means that two different orthonormal bases
%where in one case high probability is associated with complex
%basis states and low probability with simple basis states
%and {\em vice versa} have the same density matrix and hence
%von Neumann entropy, while the average Kolmogorov complexities
%are quite different.
In contrast, in the approach of \cite{Ga00}, using semicomputable
semi-density matrices, as discussed in the Introduction,
equality of ``average min-log universal density'' to the
von Neumann entropy (up to the Kolmogorov complexity
of the semicomputable density itself) 
follows simply and similarly to the classical case. 
But in this approach the interpretation of 
``min-log universal density'' in terms
of length of descriptions of one form or the other is quite problematic
(in contrast with the classical case) and we thus lose the main motivation
of quantum Kolmogorov complexity.
\end{itemize}

%It is important that the roles of $\ket{x}$ and $\ket{z}$ in
%this definition are not symmetric: $\ket{z}$ is an effectively 
%describable pure quantum state and can possibly be 
%an outcome of a measurement, but $\ket{x}$ can be every 
%one of continuously many
%pure quantum states and therefore be uncomputable (since there
%are only countably many programs and hence computable objects).
%In particular, $\ket{x}$ can be an ``outcome'' that 
%cannot be effectively prepared.

\subsection{Consistency with Classical Complexity}
Our proposal would not be useful if it were the case that for
a directly computable object the complexity is less than the
shortest program to compute that object. This would imply
that the code corresponding to the
probabilistic component in the description is possibly shorter than
the difference in program lengths for programs for an approximation
of the object and the
object itself. This would penalize definite description compared
to probabilistic description and in case of classical objects
would make quantum Kolmogorov complexity less than classical
Kolmogorov complexity. 

\begin{theorem}[Consistency]\label{theo.equiv}
Let $U$ be the reference universal quantum Turing machine
and let $\ket{x}$ be a basis vector in a directly computable orthonormal
basis ${\cal B}$, given $y$: there is
a program $p$ such that $U(p, y)= \ket{x}$.
Then $K(\ket{x} \mid y)= \min_p \{l(p): U(p, y)= \ket{x} \}$
up to $\eqa K({\cal B}\mid y) $.
\end{theorem}
\begin{proof}
Let $\ket{z}$ be such that 
\[
K (\ket{x} \mid y ) =
\min_{q} \{ l(q) + \lceil - \log \norm{\bracket{z}{x}}^2 \rceil :
U(q, y) = \ket{z} \} .
\]

Denote the program $q$ that minimizes the righthand side
by $q_{\min}$
and the program  $p$ that minimizes the expression in the statement
of the theorem by $p_{\min}$.

A {\em dovetailed} computation is a method related
to Cantor's celebrated diagonalization method: run all programs alternatingly 
in such a way that every program eventually makes progress. On
an list of programs $p_1, p_2, \ldots$ one divides
the overall computation into stages $k=1,2, \ldots$. 
In stage $k$ of the overall computation one
executes the $i$th computation step of every program $p_{k-i+1}$
for $i=1, \ldots , k$.

By running $U$ on all binary strings (candidate programs)
 simultaneously dovetailed-fashion,
one can enumerate all objects that are directly computable, given $y$,
in order of their halting programs. Assume that $U$ is also
given a $K({\cal B}\mid y)$ length program $b$ to compute 
${\cal B}$---that is, enumerate the basis
vectors in ${\cal B}$.
This way $q_{\min}$ computes
$\ket{z}$, the program $b$ computes ${\cal B}$.
Now since the vectors of ${\cal B}$ are mutually orthogonal
\[ \sum_{\ket{e} \in {\cal B}} \norm{ \bracket{z}{e}}^2 = 1 .
\] 
Since $\ket{x}$ is one of the basis vectors
we have $- \log \norm{\bracket{z}{x}}^2$ is the length of 
a prefix code (the Shannon-Fano code) to compute $\ket{x}$ from $\ket{z}$
and ${\cal B}$.
Denoting this code by $r$ we have that the concatenation $q_{\min} b r$
is a program to compute $\ket{x}$: parse it into 
$q_{\min}, b,$ and $r$ using the self-delimiting 
property of $q_{\min}$ and $b$. Use
$q_{\min}$ to compute $\ket{z}$ and use $b$ to compute ${\cal B}$, 
determine the
probabilities $\norm{\bracket{z}{e}}^2$ for all basis vectors
$\ket{e}$ in ${\cal B}$. Determine the Shannon-Fano code words
for all the basis vectors from these probabilities.
Since $r$ is the code word for $\ket{x}$ we can now
decode $\ket{x}$. Therefore,
\[ l(q_{\min} ) + \lceil - \log \norm{\bracket{z}{x}}^2 \rceil
\gea l( p_{\min}) - K({\cal B} \mid y) , \]
which was what we had to prove.
\end{proof}

\begin{corollary}\label{cor.clasquant}
\rm
On classical objects (that is, the natural numbers
or finite binary strings that are all directly computable) the 
quantum Kolmogorov complexity coincides up
to a fixed additional constant with the self-delimiting
Kolmogorov complexity since $K({\cal B} \mid n) \eqa 0$ for the standard
classical basis ${\cal B}= \{0,1\}^n$.
(We assume that the information about the dimensionality 
of the Hilbert space is given conditionally.)
\end{corollary}

\begin{remark}
\rm
Fixed additional constants are no problem since
the complexity also varies by fixed additional constants due to the choice of
reference universal Turing machine.
\end{remark}

\begin{remark}
\rm
The original plain complexity defined by Kolmogorov, \cite{LiVi97}, is
based on Turing machines where the input is delimited by distinguished markers.
A similar proof used to compare quantum Kolmogorov complexity
with the plain (not self-delimiting) Kolmogorov complexity on classical
objects shows
that they coincide, but only up to a logarithmic additive term. 
\end{remark}

\subsection{Upper Bound on Complexity}
%One way to achieve an *upper bound* of $n$ is to make $p$ a program that just
%outputs $n$ qubits on its "input tape", which specify $\ket{x}$.
%In this case, the "error term" $\log(|\bracket{y}{x}|^2)$
%is zero, so $(|P| - \log(|\bracket{y}{x}|^2)) = n$.
%\begin{corollary}
%By the existence of the identity quantum machine we have:
%For each state $\ket{x}$ consisting of $n$ qubits we have
%$C(\ket{x}) \leq n+O(1)$.
%\end{corollary}
%Another way to achieve an upper bound of $n$ with high probability is to
%make $p$ a very short program that just outputs the state 
%$\ket{ 00 \ldots 0}$.
%Here $l(p)$ can be 1, but the expected value of $ |\bracket{y}{x}|^2$ will be $1/2^n$
%(Wim can explain why). I think that with high probability 
%$|\bracket{y}{x}|^2$
%will be close to its mean, so that with high probability,
%$-\log(|\bracket{x}{y}|^2) = n$. 
%
%There are still other "hybrid" ways of getting the expected upper bound
%to be $n$, such as having $l(p) = n/2 $, and $-\log(|\bracket{y}{x}|^2) = n/2$.

A priori, in the worst case $K(\ket{x}  \mid n )$
is possibly $\infty$. We show that the worst-case has a $2n$ upper bound.

\begin{theorem}[Upper Bound]\label{lem.ubKC}
For all $n$-qubit quantum states $\ket{x}$ we
have $K(\ket{x}  \mid n)\lea 2n$.
\end{theorem}
\begin{proof}
Write $N=2^n$.
For every state $\ket{x}$ in $(2^n)$-dimensional Hilbert space
with basis vectors $\ket{e_0}, \ldots , \ket{e_{N-1}}$ we have
$\sum_{i=0}^{N-1} \norm{\bracket{e_i }{x}}^2 =1$. Hence there is an $i$
such that $\norm{\bracket{e_i }{x}}^2 \geq 1/N$. 
Let $p$ be a $\eqa K(i \mid n)$-bit program to construct a
basis state $\ket{e_i}$ given $n$.
Then $l(p) \lea n $.
Then $K ( \ket{x}  \mid n ) \leq l(p)  - \log (1/N) \lea 2n $.
\end{proof}

\begin{remark}
\rm
This upper bound is sharp since
G\'acs \cite{Ga00} has recently shown
that there are states $\ket{x}$ with $K(\ket{x}  \mid n) \gea 2n - 2 \log n$.
\end{remark}

\subsection{Computability}

In the classical case Kolmogorov complexity is not computable
but can be approximated from above by a computable process.
The non-cloning property prevents us from perfectly copying an unknown pure 
quantum state given to us \cite{WZ82,Di82}. Therefore, an approximation from
above that requires checking every output state against the
target state destroys the latter. It is possible to prepare
approximate copies from the target state, but the more copies one prepares
the less they approximate the target state
\cite{GM97}, and this deterioration appears on the surface to prevent
use in our application below. To sidestep the fragility of
the pure quantum target state, we simply require that it
is an outcome, in as many copies as we require,
in a measurement that we have available.
Another caveat with respect to item (ii) in the theorem
below is that, since the approximation algorithm in the proof
doesn't discriminate between entangled output states and
un-entangled output states, we approximate the quantum
Kolmogorov complexity by a directly computed part that 
is possibly a mixture rather than a pure state. Thus, the approximated
value may be that of quantum Kolmogorov complexity based on computations
halting with entangled
output states, which is conceivably less
than that of un-entangled outputs. 
This is the only result in this paper that depends on that distinction.

\begin{theorem}[Computability]\label{theo.appr}
Let $\ket{x}$ be the pure quantum state we want to describe.

{\rm (i)} The quantum Kolmogorov complexity $K(\ket{x})$ is not computable.

{\rm (ii)} 
If we can repeatedly execute the projection $\shar{x}{x}$
and perform a measurement with outcome $\ket{x}$, then
the quantum Kolmogorov complexity $K(\ket{x})$
can be approximated
from above by a computable process with arbitrarily
small probability of error $\alpha$ of giving a too small value.
\end{theorem}
\begin{proof}
The uncomputability follows a fortiori from the classical case.
The semicomputability follows because we have established an upper
bound on the quantum Kolmogorov complexity, and we can simply
enumerate all halting classical programs up to that length by running their
computations dovetailed fashion. The idea is as follows:
Let the target state be $\ket{x}$
of $n$ qubits. Then, $K(\ket{x} \mid n) \lea 2n $. (The 
unconditional case $K(\ket{x})$ is similar 
with $2n$ replaced by $2(n + \log n)$.)
We want to identify a program $x^*$ such that $p=x^*$ minimizes
$l(p) - \log \norm{\bracket{x}{U(p,n)}}^2$ among all candidate programs.
To identify it in the limit,
for some fixed $k$ satisfying (\ref{eq.alpha}) below 
for given $n, \alpha , \epsilon$,
repeat the computation of every halting program
$p$ with $l(p) \lea 2n$ at least $k$ times and perform the assumed
projection and measurement. For every halting program $p$ in the dovetailing
process we estimate the probability 
$q = \norm{\bracket{x}{U(p,n)}}^2$ from the fraction $m/k$:
the fraction of $m$ positive outcomes out of $k$ measurements.
The probability that the estimate $m/k$ is off from the real
value $q$ by more than
an $\epsilon q$ is given by Chernoff's bound: 
for
$0 \leq \epsilon \leq 1$,
\begin{equation}
\label{chernoff}
P ( |m- qk |  >  \epsilon qk )
\leq 2e^{ - \epsilon^2 qk /3}.
\end{equation}
This means that the probability that the deviation $|m/k - q|$
exceeds $\epsilon q$ vanishes exponentially with growing $k$.
Every candidate program $p$ satisfies
(\ref{chernoff}) with its own $q$ or $1-q$. There are $O(2^{2n})$
candidate programs $p$ and hence also $O(2^{2n})$ outcomes $U(p,n)$
with halting computations. 
%Choose $\epsilon > 0$ so that 
%if $m$ is the number of successes in the experiment
%with probability $q$, and $m'$ that for $q'$, then 
%\begin{equation}\label{eq.q}
%|\frac{m}{k}-\frac{m'}{k}| >  3 \epsilon (\frac{m}{k}+{m'}{k}). 
%\end{equation}
%Then, with probability at most $4e^{ - \epsilon^2 k /3}$ the
%two associated probabilities $q,q'$ do not satisfy
%\begin{equation}\label{eq.q}
%|q-q'| >   \epsilon (q+q') \geq  \epsilon. 
%\end{equation}
We use this estimate to upper bound the probability of error $\alpha$.
For given $k$, the probability
that {\em some} halting candidate program $p$ satisfies
$ |m- qk |  >  \epsilon qk$
is at most $\alpha$ with
\[ \alpha \leq  \sum_{U(p,n) < \infty } 2e^{ - \epsilon^2 q k /3} .\]
The probability that {\em no} halting program does so is
at least $1- \alpha$. That is, with probability
at least $1-\alpha$ we have
\[ (1- \epsilon)q  \leq \frac{m}{k} \leq (1+ \epsilon)q  \]
for every halting program $p$.
It is convenient to restrict attention to the case that all $q$'s are large. 
Without loss of generality,
if $q < \frac{1}{2}$ then consider $1-q$ instead of $q$.
Then, 
\begin{equation}\label{eq.alpha}
 \log \alpha \lea 2n- (\epsilon^2 k \log e )/ 6 .
\end{equation}

The approximation algorithm is as follows:

{\bf Step 0:} Set the required degree of approximation $\epsilon < 1/2$
and the number of trials $k$ to achieve the required probability of error $\alpha$.

{\bf Step 1:} Dovetail the running of all candidate programs until the
next halting program is enumerated.
Repeat the computation of the new halting program $k$ times

{\bf Step 2:} If there is more than one program $p$ that achieves the
current minimum, then choose the program with the least length
(and hence the least number of successfull observations).
If $p$ is the selected program with $m$ successes out of $k$ trials
then set the current approximation of $K(\ket{x})$ to
\[l(p) - \log \frac{m}{(1+\epsilon)k} .\]
This exceeds the proper value
of the approximation based on the real $q$ instead
of $m/k$ by at most 1 bit for all $\epsilon < 1$.

{\bf Step 3:} Goto {\bf Step 1}.
\end{proof}

\subsection{Incompressibility}

\begin{definition}\label{def.pqscomp}
\rm
A pure quantum state $\ket{x}$ is {\em computable} if 
$K(\ket{x}) < \infty$. Hence all finite-dimensional pure
quantum states are computable. 
We call a pure quantum state {\em directly
computable} if there is a program $p$ such that
$U(p)= \ket{x}$.
\end{definition}

We have shown that quantum Kolmogorov complexity coincides
with classical Kolmogorov complexity on classical objects in
Theorem~\ref{theo.equiv}. In the proof we demonstrated in fact
that the quantum Kolmogorov complexity is the length of the 
classical program that directly computes the classical objects.
By the standard counting argument, Section~\ref{sect.kc},
the standard orthonormal basis---consisting of all $n$-bit strings---of
 the $(2^n)$-dimensional
Hilbert space ${\cal H}_N$ ($N=2^n$) has
at least $2^n (1-2^{-c})$ basis vectors $\ket{e_i}$
that satisfy $K(\ket{e_i} \mid n) \geq n-c$. 
But what about nonclassical orthonormal bases? They may not satisfy
the standard counting argument. Since there are continuously
many pure quantum states and the range of quantum Kolmogorov
complexity has only countably many values, there are integer values
that are the Kolmogorov complexities of continuously many pure quantum states.

In particular, since the quantum Kolmogorov complexity
of an $n$-qubit state is $\lea 2n$, the set of directly
computable pure $n$-qubit states has cardinality 
$A \leq 2^{2n+O(1)}$. They divide the set of unit vectors in ${\cal H}_N$,
the surface of the $N$-dimensional ball with unit radius in Hilbert space,
into $A$-many $N-1$ dimensional connected surfaces, called {\em patches},
 each consisting  
of one directly computable pure  $n$-qubit state $\ket{x}$ together
with those pure $n$-qubit states $\ket{y}$ of which $\ket{x}$
is the directly computed part (Definition~\ref{def.qkc}). In every patch
all $\ket{y}$ with the same
$\norm{\bracket{x}{y}}$ have both the same complexity and the same
directly computed part, and for every fixed patch and every fixed value of 
approximation part occurring in the patch, there are continuously
many $\ket{y}$ with identical directly computed parts and approximation
parts.
A priori it is possible that this is the case for two distinct basis vectors
in a nonclassical orthonormal bases, which implies that the standard
counting argument cannot be used to show the incompressibility of
basis vecors of nonclassical orthonormal bases.
\begin{lemma}\label{lem.lowb}
There is a particular  (possibly nonclassical) 
orthonormal basis of the $(2^n)$-dimensional
Hilbert space ${\cal H}_N$, computed from the directly computed pure
quantum states, such
that at least $2^n (1-2^{-c})$ basis vectors $\ket{e_i}$ 
satisfy $K(\ket{e_i} \mid n) \geq n-c$.
\end{lemma}
\begin{proof}
Every orthonormal basis of ${\cal H}_N$ 
has $2^n$ basis vectors and there are at most
$m  \leq \sum_{i=0}^{n-c-1} 2^i = 2^{n-c}-1$ programs of length less than
$n-c$.  Hence there are at most $m$ programs of length $< n-c$
available to approximate the basis vectors.
We construct an orthonormal basis satisfying the lemma:
The set of directly computed pure quantum states 
$\ket{x_0}, \ldots , \ket{x_{m-1}}$
span an $m'$-dimensional subspace ${\cal A}$ with $m' \leq m$
in the $(2^n)$-dimensional Hilbert space ${\cal H}_N$ such
that ${\cal H}_N = {\cal A} \oplus {\cal A}^{\perp}$.
Here ${\cal A}^{\perp}$ is a $(2^n - m')$-dimensional
subspace of ${\cal H}_N$ such that every vector in it is
perpendicular to every vector in ${\cal A}$.  
We can write every element $\ket{x} \in {\cal H}_N$ as 
\[
\sum_{i=0}^{m'-1} \alpha_i \ket{a_i}+ \sum_{i=0}^{2^n-m'-1} \beta_i \ket{b_i}
\]
where the $\ket{a_i}$'s form an orthonormal basis
of ${\cal A}$ and the  $\ket{b_i}$'s form an
orthonormal basis of $ {\cal A}^{\perp}$ so that
the $\ket{a_i}$'s and $\ket{b_i}$'s form an orthonormal basis $K$
for ${\cal H}_N$. For every state
 $\ket{x_j} \in {\cal A}$, directly computed by a program
$x^*_j$, given $n$, and basis vector 
$\ket{b_i} \in {\cal A}^{\perp}$ we have
$\norm{\bracket{x_j}{b_i} }^2 = 0$.
Therefore, 
$K(\ket{b_i} \mid n) \gea l(x^*_j) - \log \norm{ \bracket{x_j}{b_i} }^2 
= \infty
> n-c$
($0 \leq j < m$, $0 \leq i < 2^n - m'$).
This proves the lemma.
\end{proof}

\begin{theorem}[Incompressibility]
The uniform probability 
$\Pr\{\ket{x}: l(\ket{x})=n, \; \; K(\ket{x} \mid n) \geq n-c \} \geq 1-1/2^c$.
\end{theorem}

\begin{proof}
The theorem follows immediately
from a generalization of Lemma~\ref{lem.lowb} to arbitrary
orthonormal bases:
\begin{claim}
Every orthonormal basis $\ket{e_0}, \dots ,
\ket{e_{2^n-1}}$ of the $(2^n)$-dimensional
Hilbert space ${\cal H}_N$  has
at least $2^n (1-2^{-c})$ basis vectors $\ket{e_i}$ 
that satisfy $K(\ket{e_i} \mid n) \geq n-c$.
\end{claim}
\begin{proof}
Use the notation of the proof of Lemma~\ref{lem.lowb}.
Let $A$ be a set initially containing the programs of length less
than $n-c$, and let $B$ be a set initially containing the set of 
basis vectors $\ket{e_i}$ with $K(\ket{e_i} \mid n) < n-c$.
Assume to the contrary that $|B| >2^{n-c}$.
Then at least two of them, say $\ket{e_0}$
and $\ket{e_1}$ and some
pure quantum state $\ket{x}$ directly computed from a $<(n-c)$-length program
satisfy 
\begin{equation}\label{eq.ex}
K(\ket{e_i} \mid n) = K(\ket{x} \mid n) + \lceil 
- \log \norm{\bracket{e_i}{x}}^2 \rceil ,
\end{equation}
with $\ket{x}$ being the directly computed part of both $\ket{e_i}$, $i=0,1$.
This means that $K(\ket{x} \mid n)<n-c-1$ since not both
$\ket{e_0}$ and $\ket{e_1}$ can be equal to $\ket{x}$.
Hence for every directly computed pure quantum state of complexity
$n-c-1$ there is at most one basis state, say $\ket{e}$, of the same complexity
(in fact only if that basis state is identical with the directly
computed state.)
Now eliminate every directly computed pure quantum state $\ket{x}$ of 
complexity $n-c-1$ from the set $A$,
and the basis state $\ket{e}$ as above (if it exists) from $B$.
We are now
left with $|B| > 2^{n-c} -1$ basis states of which the directly
computed parts are included in $A$ with
$|A| \leq 2^{n-c-1}-1$ with every element in $A$
of complexity $\leq n-c-2$.
Repeating the same argument we end up with $|A|>1$ basis vectors
of which the
directly computed parts are elements of the empty set $B$,
which is impossible.
\end{proof}
\end{proof}

\begin{example}
\rm
It may be instructive to check the behavior
of the approximation part $- \log  \norm{ \bracket{x}{z} }^2$
in Definition~\ref{def.qkc} on a nontrivial example.
Let $x$ be a random classical string with $K(x) \geq l(x)$ 
 and let $y$ be a string obtained from $x$
by complementing one bit, say in position $j$. 
It is known (Exercise 2.2.8 in \cite{LiVi97}
due to I. Csisz\'ar)
that for every such $x$ of length $n$ there is such a $y$ with complexity 
$K(y \mid n) \eqa n - \log n $. 
Since $K(x \mid n) \lea K(y \mid n) + K(j \mid n)$ 
we have $K(j \mid n ) \gea \log n$ (and, since $j \leq n$ we also
have $K(j \mid n ) \lea \log n$).
Now let $\ket{z}$ be a pure quantum state which has
classical bits except the difference qubit between $x$ and $y$ that has
equal probabilities of being observed as ``1'' and as ``0.''
We can prepare $\ket{z}$ by giving $y$ and the position of the
difference qubit (in $\log n$ bits) 
and therefore $K(\ket{z} \mid n) \lea n $. 

From
$\ket{z}$ we have probability $\frac{1}{2}$ of obtaining $x$
by observing the difference qubit,
it follows $K(x \mid n) \lea K(\ket{z}  \mid n,j)$, and,
since $K(\ket{z} \mid n) \gea K(\ket{z}  \mid n,j)$,
we have $K(\ket{z} \mid n) \gea n $.

From $\ket{z}$ we also have probability $\frac{1}{2}$ of obtaining $y$
by observing the difference qubit which yields that
$K(y \mid n) \lea K(\ket{z} \mid n,j)$. Since also
$K(\ket{z} \mid n) \gea K(\ket{z} \mid n,j) 
\gea K(\ket{z} \mid n) - K(j   \mid n) \eqa 
K(\ket{z} \mid n) \eqa n - \log n$,
we find $n - \log n \lea K(y \mid n) \lea n$. This is the strongest
conclusion we can draw about $y$ from the fact that it
is the result of observing one qubit 
of a high-complexity $\ket{z}$ constructed as above.
Viz., if we flip an $i$th bit of $x$  with complexity
$K(i   \mid n) \eqa \log n$, this will not necessarily
result in a string of complexity
$\eqa n - \log n$ (take for example $i=j/2$ with $j$ as above).
\end{example}

\begin{remark}
\rm
Theorem~\ref{lem.ubKC} states an upper bound of $2n$ on $K(\ket{x}  \mid n)$.
This leaves a relatively large gap with the lower bound of $n$ 
established here. But, as stated earlier, G\'acs \cite{Ga00} has shown
that there are states $\ket{x}$ with $K(\ket{x}  \mid n) \gea 2n - 2 \log n$;
in fact, most states satisfy this. The proof appears to support
about the same incompressiblity results as
in this section, with $n$ replaced by $2n-2 \log n$.
The proof goes by analyzing coverings of
the $(2^n)$-dimensional ball of unit radius, as in \cite{CS98}.
\end{remark}

\subsection{Multiple Copies}
For classical complexity we have $K(x,x) \eqa K(x)$, since a 
classical program to compute $x$ can be used twice; indeed,
it can be used many times. In the quantum world things
are not so easy: the no-cloning property mentioned earlier,
see \cite{WZ82,Di82} or the textbooks \cite{Pe95,NC00}, prevent
cloning an unknown pure state $\ket{x}$ perfectly to obtain $\ket{x} \ket{x}$:
that is, $K(\ket{x}) < K(\ket{x} \ket{x}) \lea 2K(\ket{x})$. 
There is a considerable literature
on the possibility of approximate cloning
to obtain $m$ imperfect copies from an unknown pure state, see
for example \cite{GM97}. Generally speaking, the more qubits
are involved in the original copy and the more clones
one wants to obtain, the more the fidelity of the obtained clones
deteriorates with respect to the original copy. This stands to reason since
high fidelity cloning would enable both superluminal signal transmission 
\cite{Gi98}
and extracting essentially unbounded information 
concerning the probability amplitude
from the original qubits. The approximate cloning possibility
suggests that in our setting
the approximation penalty induced by the second---fidelity---term of 
Definition~\ref{def.qkc} may be lenient insofar that the
complexity of multiple copies increases sublinearly with the number of copies. 
Even apart from this, the $m$-fold
tensor product $\ket{x}^{\otimes m}$ of $\ket{x}$ with itself
lives in a small-dimensional symmetric subspace 
with the result that $K(\ket{x}^{\otimes m})$  can be considerably
below $m K(\ket{x} )$. This effect was first noticed in the context
of qubit complexity \cite{BDL00}, and it similarly holds for
the $Kg$ and $KG$ complexities in \cite{Ga00}. 
Define $K^+(\ket{x}^{\otimes m}) = \max \{ K(\ket{x}^{\otimes m}):
\ket{x} \; \; \mbox{\rm  is a pure $n$-qubit quantum state} \}$ and
write $N=2^n$.
The following theorem states that the $m$-fold copy of 
{\em every} $n$-qubit
pure quantum state has complexity at most about $4 \log {{m+N-1} \choose m}$,
and {\em there is} a pure quantum state for
which the complexity of the $m$-fold copy achieves $ \log {{m+N-1} \choose m}$.

\begin{theorem}[Multiples]\label{theo.nocloning}
Assume the above terminology.
\begin{eqnarray*}
 \log {{m+N-1} \choose m} & \leq &
 K^+(\ket{x}^{\otimes m}) \\
& \lea &
4 \left[ K(m) +  \log {{m+N-1} \choose m} \right] \\
&& +
2 \log \left[ K(m) +  \log {{m+N-1} \choose m}  \right]. 
\end{eqnarray*}
\end{theorem}
\begin{proof}
Recall the $Kg$ and $KG$ complexities of pure quantum states
\cite{Ga00}
mentioned in the Introduction. 
Denote by $Kg^+ (\ket{x}^{\otimes m})$ and $KG^+ (\ket{x}^{\otimes m})$
the maximal values of  $Kg (\ket{x}^{\otimes m})$ and $KG (\ket{x}^{\otimes m})$
over all $n$ qubit states $\ket{x}$, respectively.
All of the following was shown in \cite{Ga00} (the notation as above and
$\ket{y}$ an arbitrary state, for example $\ket{x}^{\otimes m}$):
\begin{eqnarray*}
&&KG^+ (\ket{x}^{\otimes m}) \lea K(m) +  \log {{m+N-1} \choose m} 
\\
&&Kg^+ (\ket{x}^{\otimes m}) \geq  \log {{m+N-1} \choose m}  \\
&& Kg ( \ket{y}) \leq KG(\ket{y}) \\
&&Kg(\ket{y}) \lea K(\ket{y}) \lea 4 Kg (\ket{y})+ 2 \log Kg(\ket{y}) \\
\end{eqnarray*}
Combining these inequalities gives the theorem.
\end{proof}

The theorem gives a measure of how ``clonable'' {\em individual}
$n$-qubit pure quantum 
states are---rather than indicate the {\em average} success of a 
fixed cloning
algorithm for all $n$-qubit pure quantum states,
as in the approximate and probabilistic cloning algorithms
referred to above. In particular it gives
an upper bound on the non-clonability of every individual pure quantum
state, and moreover it tells us that there 
exist individual pure quantum states that are quite non-clonable.
One can view this as an application of quantum Kolmogorov complexity. 
The difference $K^+(\ket{x}^{\otimes m}) - K^+(\ket{x}) $
expresses the amount of extra information required
for $m$ copies of $\ket{x}$ over that of one copy---in our
particular meaning of (\ref{eq.qkc}).

\subsection{Conditional Complexity and Cloning}
In Definition~\ref{def.qkc} the conditional 
complexity $K(\ket{x} \mid y)$
is the minimum sum of the length of a classical program to compute
$\ket{z}$ plus the negative logarithm of the probability of outcome
$\ket{x}$ when executing projection $\shar{x}{x}$ on $\ket{z}$
and measuring, given $y$ as input on an auxiliary input
tape. In case $y$ is a classical object, a finite binary string,
there is no problem with this definition. The situation
is more complicated if instead of a classical `$y$' we
consider the pure quantum
state $\ket{y}$ as input on an auxiliary ``quantum'' input tape. 
In the quantum situation the notion of inputs consisting
of pure quantum states is subject to very special rules.

Firstly, if we are given an unknown pure quantum state $\ket{y}$ as
input it can be used only once, that is, it is irrevocably consumed
and lost in the computation. It cannot be perfectly copied or cloned without
destroying the original as discussed above.
This means that there is
a profound difference between representing a directly computable pure quantum
state on the auxiliary tape as a classical program 
or giving it literally. Given as a 
classical program we can prepare and use arbitrarily many copies of it.
Given as an (unknown) pure quantum state in superposition it can be
used as perfect input to a computation only once.
Thus, the manner in which the conditional information
is provided may make a great difference. A classical program for
computing a directly computable quantum state carries {\em more information}
than the directly computable quantum state itself---much like a
shortest program for a classical object carries more information than the
object itself. In the latter case it consists in partial information
about the halting problem. In the quantum case of a directly
computable pure state we have the additional
information that the state is directly computable {\em and} 
in case of a shortest classical program additional information
about the halting problem. 
Thus, for classical objects $x$
we have $K(x^m \mid x) \eqa K(m)$ in contrast to:
\begin{theorem}[Cloning]
For every pure quantum state $\ket{x}$ and every $m$,
we have:
\begin{equation}\label{eq.double1}
K(\ket{x}^{\otimes m}  \mid  \ket{x}) \lea K(\ket{x}^{\otimes m-1}).
\end{equation}
Moreover, for every $n$ there exists an $n$-qubit
 pure quantum state $\ket{x}$, such that for every $m$,
we have:
\begin{equation}\label{eq.double2}
K(\ket{x}^{\otimes m}  \mid  \ket{x}) \gea \frac{1}{4}K(\ket{x}^{\otimes m-1}).
\end{equation}
\end{theorem}

\begin{proof}
(\ref{eq.double1}) is obvious. 
(\ref{eq.double2}) follows from Theorem~\ref{theo.nocloning}.
\end{proof}

This
holds even if $\ket{x}$ is directly computable but is
given in the conditional in the form of an unknown pure quantum state. 
The lemma quantifies the ``no-cloning'' property of an individual
pure quantum satte $\ket{x}$: Given $\ket{x}$ and the
task to obtain $m$ copies of $\ket{x}$, we require
at least $\frac{1}{4}$th of the information to 
optain $m-1$ copies of $\ket{x}$---everything in the sense
of quantum Kolmogorov complexity (\ref{eq.qkc}).
However,
if $\ket{x}$ is directly computable and
the conditional is a classical program to compute
this directly computable state, then 
that program can be used over and over again, just like in the
case of classical objects:
\begin{lemma}
For every directly computable pure quantum state $\ket{x}$
computed by a classical program $p$, and every $m$, 
\begin{equation}\label{eq.double3}
K(\ket{x}^{\otimes m}  \mid  p,m) \eqa 0 .
\end{equation}
\end{lemma}

\subsection{Sub-additivity}
Let $N=2^n$ and $M=2^m$. Recall the
following notation: If $\ket{x}$ is a pure quantum state in $(2^n)$-dimensional
Hilbert space of $l(\ket{x})=n$ qubits,
 and $\ket{y}$ is a pure quantum state in $(2^m)$-dimensional
Hilbert space of $l(\ket{y})=m$ qubits, then $\ket{x} \otimes \ket{y} = \ket{x} \ket{y} =
\ket{x,y}$ is a pure quantum state in the $NM$-dimensional Hilbert space
consisting of the tensor product of the two initial spaces consisting
of $l(\ket{x,y})=n+m$ qubits.

In the classical Kolmogorov complexity
case we have $K(x) \lea K(x,y) \lea K(x|y)+K(y)$
for every pair of {\em individual} finite binary strings $x$ and $y$ 
(the analog of the similar familiar relation that holds among
entropies---a stochastic notion---in
Shannon's information theory). 
The second inequality is the {\em sub-additivity} property
of classical Kolmogorov complexity. Obviously, in the quantum setting
also $K(\ket{x,y}) \gea K(\ket{x})$ for every pair of 
individual pure quantum states
$\ket{x}, \ket{y}$.
Below we shall show that the sub-additive property does {\em not}
hold for quantum Kolmogorov complexity. But in the restricted
case of 
directly computable pure
quantum states in simple orthonormal bases quantum
Kolmogorov complexity  {\em is} sub-additive,
just like classical Kolmogorov complexity:
\begin{lemma}\label{lem.additive}
For directly computable $\ket{x}, \ket{y}$ both of which
belong to (possibly different) orthonormal bases of
Kolmogorov complexity $O(1)$ we have
\[ K(\ket{x}, \ket{y} ) \lea K(\ket{x} \mid \ket{y}) + K(\ket{y}) \]
up to an additive constant term.
\end{lemma}
\begin{proof}
By Theorem~\ref{theo.equiv} we there is a program $p_y$ to compute $\ket{y}$
with $l(p)= K(\ket{y})$ and, by a similar argument as used in the proof
of Theorem~\ref{theo.equiv}, a program
$p_{y \rightarrow x}$ to compute $\ket{x}$ from $\ket{y}$ 
with $l(p_{y \rightarrow x}) = K(\ket{x} \mid \ket{y})$ up
to additive constants. Use $p_y$ to
construct two copies of $\ket{y}$ and $p_{y \rightarrow x}$ to construct
$\ket{x}$ from one of the copies of $\ket{y}$.
The separation between
these concatenated binary programs is taken care of
by the self-delimiting property
of the subprograms. An additional constant term 
takes care of the couple of $O(1)$-bit
programs that are required.
\end{proof}

In the classical case we have equality in the lemma (up
to an additive logarithmic term). 
The proof of the remaining inequality, as given in the classical case,
see \cite{LiVi97},
doesn't hold for the quantum case. It would require
a decision procedure that establishes equality between 
two pure quantum states without error. 
It is unknown to the author whether some approximate decision rule
would give some result along the required lines.
We additionally note:
\begin{lemma}
For all directly computable pure states $\ket{x}$ and $\ket{y}$ we have
$K(\ket{x}, \ket{y}) \leq K(\ket{y}) - \log \norm{ \bracket{x}{y}}^2$
up to an additive logarithmic term.
\end{lemma}
\begin{proof}
$ K(\ket{x} \mid \ket{y}) \leq  - \log \norm{ \bracket{x}{y}}^2$ by the proof
of Theorem~\ref{theo.equiv}.
Then, the lemma follows by Lemma~\ref{lem.additive}.
%(Tentative) Suppose $U(p)=\ket{z}, U(q)=\ket{t}$ and
%$K(\ket{x}) = l(p) - \log \norm{ \bracket{x}{z}}^2$
%$K(\ket{y}) = l(q) - \log \norm{ \bracket{y}{z}}^2$
\end{proof}

In  contrast, quantum Kolmogorov complexity of arbitrary individual
pure quantum states dramatically {\em fails} to be sub-additive:
\begin{theorem}[Sub-additivity]
There are pure quantum states $\ket{x}$, $\ket{y}$ of every length $n$
such that $K(\ket{x,y}) > K(\ket{x}) > K(\ket{x} \mid \ket{y}) + K(\ket{y})$.
\end{theorem}
\begin{proof}
Only the second inequality is non-obvious.
Let $\ket{y} = \frac{1}{\sqrt{2}} ( \ket{00 \ldots 0} + \ket{x} )$
and let $x$ be a maximally complex classical $n$-bit state.
Then, $- \log \norm{ \bracket{y}{x}}^2 = 1$. Hence the $O(1)$-bit program
approximating $\ket{x}$ by observing input $\ket{y}$,
and outputting the resulting outcome,
demonstrates $K(\ket{x} \mid \ket{y}) \eqa 0$.
Furthermore, $\ket{y}$ is approximated by $\ket{00 \ldots 0}$
with $- \log \norm{ \bracket{00 \ldots 0}{y}}^2 = 1$. Thus, 
$K(\ket{y}) \lea \log n + 2 \log \log n$ (the $\log$-term is
due to the specification of the length of $\ket{00 \ldots 0}$,
and the $\log \log$ term
is due to the requirement of self-delimiting
coding). The lemma follows since
$K(\ket{x}) \gea n$.
\end{proof}

%If an unknown state $\ket{y}$ is given as auxiliary input (conditional)
%then the no-cloning theorem (\cite{NC00})
%of quantum computing says it can be used
%only {\em once}. Even so, since $K(\ket{y})$ consists of a directly
%computed part and an approximation part, we cannot conclude 
%$ K(\ket{x} \mid \ket{y} ) \gea K(\ket{x}) - K(\ket{y})$
%as in the classical case.
Note that the witness states in the proof have
$K(\ket{x} \mid \ket{y}) + K(\ket{y})\lea \log n$. If we add the length
$n$ in the qubit state in the conditional, then the upper bound reduces to
$\eqa 0$, while the lefthand-side in the lemma stays $\gea n$. 
In the light of Theorem~\ref{theo.equiv} (with $n$ substituted 
in the conditional)
this result indicates that
state $\ket{y}$ in the proof, although obviously
directly computable, is not directly computable as an
element from an orthonormal basis of low complexity. 
Every orthonormal basis ${\cal B}$,
of which $\ket{y}$ is
a basis element, has complexity $K({\cal B} |n) \gea n - K(\ket{y}|n) \eqa n$.

The ``no-cloning''  or ``approximate cloning'' theorems in
\cite{WZ82,Di82,GM97,Gi98,NC00,Pe95} essentially show the following:
Perfect cloning
is only possible if we measure according to an orthonormal basis
of which one of the basis elements is the pure quantum state to
be measured. Then, the measured pure quantum state can be reproduced at will.
Approximate cloning considers how to optimize measurements
so that for a random pure quantum state (possibly from a
restricted set) the reproduced clone has on average optimal fidility
with the original.  Here we see that while 
the complexity $K(\ket{y}|n)$ of the original state 
$\ket{y}$ in the proof above is $\eqa 0$, the complexity of an
orthonormal bases of which it is a basis element can be (and usually
is in view of the incompressibility theorems) at least $n$ for uniform
at random chosen states $\ket{x}$---or every other complexity
in between 0 and $n$ by choice of $\ket{x}$. 
This gives a rigorous quantification
of the quantum cloning fact that if we
have full information to reproduce the basis of 
which the unknown {\em individual}
pure quantum state $\ket{y}$ is a basis element,
then the quantum Kolmogorov complexity of that element is about zero---that
is, we can reproduce it at will.

It is easy to see that for the general case of pure states,
an alternative demonstration of why
the sub-additivity property fails, can be given
by way of the ``non-cloning'' property of Theorem~\ref{theo.nocloning}.
\begin{lemma}
There are infinitely many $m$ and $n$
such that there are pure $n$-qubit states $\ket{x}$ for which
\[   K(\ket{x}^{\otimes m} ) >
K(\ket{x}^{\otimes m/2} \mid  
\ket{x}^{\otimes m/2}) + K(\ket{x}^{\otimes m/2}),
\]
where ``$>$'' is meant in the sense of ``$\not\lea$''.
\end{lemma}
\begin{proof}
With $N=2^n$ we have
 \footnote{
Use the following formula (\cite{LiVi97}, p. 10),
\[ \log {a \choose b} = b \log \frac{a}{b} + (a-b) \log \frac{a}{a-b}
+ \frac{1}{2} \log \frac{a}{b(a-b)} + O(1) .\]
}
\[
 \log {{k+N-1} \choose k} \rightarrow k (n - \log k + \log e ) 
- \frac{1}{2} \log k + O(1),
\]
for $n \rightarrow \infty$ with $k$ fixed.
Substitution in Theorem~\ref{theo.nocloning} shows that
there exists a state $\ket{x}$ such that
(up to logarithmic additive terms)
$K(\ket{x}^{\otimes k} ) \geq kn$ and 
$K(\ket{x}^{\otimes k/8}) \leq \frac{1}{2}k n$.
So writing (again up to logarithmic additive terms)
\begin{eqnarray*}
K(\ket{x}^{\otimes k} ) & \lea &
K(\ket{x}^{\otimes k/2} \mid  
\ket{x}^{\otimes k/2}) + K(\ket{x}^{\otimes k/2}) \\
& \eqa&
K(\ket{x}^{\otimes k/2}) \\
&\lea & K(\ket{x}^{\otimes k/4} \mid  
\ket{x}^{\otimes k/4}) + K(\ket{x}^{\otimes k/4})  \\
& \eqa &
K(\ket{x}^{\otimes k/4}) \\
& \lea &
K(\ket{x}^{\otimes k/8} \mid  
\ket{x}^{\otimes k/8}) + K(\ket{x}^{\otimes k/8}) \\
& \eqa &
K(\ket{x}^{\otimes k/8}),
\end{eqnarray*}
we obtain $kn \leq \frac{1}{2} kn$, up to an additive logarithmic
term, which, with $k,n > 0$, can only hold for $k \eqa n \eqa 0$.
Hence, for large enough $k$ and $n$,
 one of the $\lea$ inequalities in the above chain must be false.

\end{proof}

\appendix
\section{Appendix: Classical Kolmogorov Complexity}
\label{sect.kc}
\label{app.A}
It is useful to summarize the relevant parts and definitions
of classical Kolmogorov complexity; see also
\cite{ViLi00}, and the textbook
\cite{LiVi97}.
The Kolmogorov complexity \cite{Ko65} of a finite object $x$
is the length of the
shortest effective binary description of $x$.
Let $x,y,z \in {\cal N}$, where
${\cal N}$ denotes the natural
numbers and we identify
${\cal N}$ and $\{0,1\}^*$ according to the
correspondence
\[(0, \epsilon ), (1,0), (2,1), (3,00), (4,01), \ldots \]
Here $\epsilon$ denotes the {\em empty word} `' with no letters.
The {\em length} $l(x)$ of $x$ is the number of bits
in the binary string $x$. For example,
$l(010)=3$ and $l(\epsilon)=0$.

The emphasis is on binary sequences only for convenience;
observations in every finite or countably infinite
 alphabet can be so encoded in a way
that is `theory neutral'.

A binary string $x$
is a {\em proper prefix} of a binary string $y$
if we can write $x=yz$ for $z \neq \epsilon$.
 A set $\{x,y, \ldots \} \subseteq \{0,1\}^*$
is {\em prefix-free} if for every pair of distinct
elements in the set neither is a proper prefix of the other.
A prefix-free set is also called a {\em prefix code}.
Each binary string $x=x_1 x_2 \ldots x_n$ has a
special type of prefix code, called a
{\em self-delimiting code},
\[ \bar x =1x_1x_1x_2x_2 \ldots x_n \neg x_n ,\]
where
$\neg x_n=0$ if $x_n=1$ and $\neg x_n=1$ otherwise. This takes
care of all strings of length $n \geq 1$. The empty string
$\epsilon$ is encoded by $\bar \epsilon = 0$. This code
is self-delimiting because we can determine where the
code word $\bar x$ ends by reading it from left to right without
backing up. Using this code we define
the standard self-delimiting code for $x$ to be
$x'=\overline{l(x)}x$. It is easy to check that
$l(\bar x ) = 2 n+1$ and $l(x')=n+2 \log n+1$.

Let $\langle \cdot ,\cdot \rangle$ be a standard one-one mapping
from ${\cal N} \times {\cal N}$
to ${\cal N}$, for technical reasons choosen such that
$l(\langle x ,y \rangle) = l(y)+O(l(x))$.
An example is $\langle x ,y \rangle = \bar {l(x)}xy$.
This can be iterated to
$\langle  \langle \cdot , \cdot \rangle , \cdot \rangle$.

Let $T_1 ,T_2 , \ldots$ be a standard enumeration
of all Turing machines, and let $\phi_1 , \phi_2 , \ldots$
be the enumeration of corresponding functions
which are computed by the respective Turing machines.
That is, $T_i$ computes $\phi_i$.
These functions are the {\em partial recursive} functions
or {\em computable} functions. 
The {\em conditional complexity} of $x$ given $y$
with respect to a Turing machine $T$ is
\[C_T(x|y) = \min_{p \in \{0,1\}^*} \{l(p): T (\langle p,y\rangle)=x \}. \]
The unconditional Kolmogorov complexity of $x$ with respect to $T$ is defined
by $C(x) = C(x| \epsilon )$.
Choose a universal Turing machine
$U$ that expresses its universality in the following manner:
\[U(\langle \langle i,p \rangle ,y \rangle ) =
T_i (\langle p,y\rangle) \]
 for all $i$ and $\langle p,y\rangle$.

\begin{theorem}[Invariance]
There is a universal Turing machine
$U$, such that for all machines $T$,
there is a constant $c_T$ (the length of  a self-delimiting
description of the index of $T$ in the enumeration),
such that for all $x$  and $y$ we have:
\[
C_U (x \mid y) \leq C_T (x\mid y) + c_T.
\]
\end{theorem}

For %
\it every %
\rm pair $U ,  U ' $
of universal Turing machines for which the theorem holds,
there
is a fixed constant $c_{{U} , U' }$,
depending only on $U$ and $U'$, such
that for all $x, y$ we have:
\[
|C_{U} (x \mid  y) - C_{U'} (x \mid y)|   \leq   c_{U , U ' } .
\]
To see this, substitute $U'$ for $T$ in the theorem,
and, conversely, substitute $U'$ for $U$ and $U$ for $T$ in the
theorem,
and combine the two resulting
inequalities.
While the complexities according to $U$ and $U'$
are not exactly equal, they are %
\it equal up to a fixed constant %
\rm for all $x$ and $y$.
Therefore, one or the other fixed choice of reference universal machine $U$
yields resulting complexities that are in a fixed constant
enveloppe from each other for all arguments.

\begin{definition}\label{def.KolmC}\label{def.KC}
We fix $U$ as our {\em reference universal computer} and define
the {\em conditional Kolmogorov complexity} of $x$ given $y$
by
\[C(x|y) = \min_{p \in \{0,1\}^*} \{l(p): U (\langle p,y\rangle)=x \}. \]
The unconditional Kolmogorov complexity of $x$ is defined
by $C(x) = C(x| \epsilon )$.
\end{definition}

The Kolmogorov complexity
$C(x)$ of $x$ is the length of the shortest binary program
from which $x$ is computed: 
Though defined in terms of a
particular machine model, the Kolmogorov complexity
is machine-independent up to an additive
constant
 and acquires an asymptotically universal and absolute character
through Church's thesis, from the ability of universal machines to
simulate one another and execute every effective process.
  The Kolmogorov complexity of an object can be viewed as an absolute
and objective quantification of the amount of information in it.
   This leads to a theory of {\em absolute} information {\em contents}
of {\em individual} objects in contrast to classic information theory
which deals with {\em average} information {\em to communicate}
objects produced by a {\em random source} \cite{LiVi97}.

{\bf Incompressibility:}
Since there is a Turing machine, say $T_i$, that computes the identity
function $T_i (x|y) \equiv x$ for all $y$,
 it follows that $C(x|y)  \leq l(x) + c$ for
fixed $c \leq  2 \log i +1$ and all $x$.
%%\footnote{``$2 \log i$'' and not ``$\log i$'' since we need to encode $i$
%%in such a way that $U$ can determine the end of the encoding. One way to do
%%that is to use the code $1^l(l(i))0l(i)i$ which has length $2l(l(i))+l(i)+1
%%< 2 \log i$ bits.
%%}

It is easy to see that there are also strings that can be described
by programs much shorter than themselves. For instance, the
function defined by $f(1) = 2$ and $f(i) = 2^{f(i-1)}$
for $i>1$ grows very fast, $f(k)$ is a ``stack'' of $k$ twos.
Yet for every $k$ it is clear that $f(k)$
has complexity at most $\eqa C(k)$.
What about incompressibility?
For every $n$ there are $2^n$ binary
strings of length $n$, but only
$\sum_{i=0}^{n-1} 2^i = 2^n -1$ descriptions in binary string
format of length less than $n$.
Therefore, there is at least one binary string
$x$ of length $n$ such that $C(x)   \geq   n$.
We call such strings $incompressible$. The same argument holds
for conditional complexity: since for every length $n$
there are at most $2^n-1$ binary programs of length $<n$,
for every binary string $y$
there is a binary string $x$ of length $n$ such that
$C(x| y)   \geq   n$.
``Randomness deficiency''
measures how far the object falls short of the maximum possible
Kolmogorov complexity.
For every constant $\delta$ we say a string $x$ is
has
\it randomness deficiency
\rm
at most $\delta$
if $C(x)   \geq   l(x) -\delta$.
Strings that are incompressible (say, with small randomness deficiency)
are patternless,
since a pattern could be used to reduce
the description length. Intuitively, we
think of
such patternless sequences
as being random, and we
use ``random sequence''
synonymously with ``incompressible sequence.''
(It is possible to give a rigorous
formalization of the intuitive notion
of a random sequence as a sequence that passes all
effective tests for randomness, see for example \cite{LiVi97}.)

Since there are few short programs, there can
be only few objects of low complexity:
the number
of strings of length $n$, that have randomness deficiency at most $\delta$,
is at least $2^n - 2^{n-\delta} +1$. Hence
there is at least one string of length $n$ with randomness deficiency 0,
at least one-half of all strings of length $n$ have randomness deficiency 1,
at least three-fourths  of all strings
of length $n$ have randomness deficiency 2, and
at least the $(1- 1/2^{\delta} )$th part
of all $2^n$ strings of length $n$ have randomness deficiency at most $\delta$.
\begin{lemma}
\label{C2}
Let $\delta$ be a positive integer.
For every fixed $y$, every
set $S$ of cardinality $m$ has at least $m(1 - 2^{-\delta} ) + 1$
elements $x$ with $C(x| y)   \geq   \lfloor \log m \rfloor  - \delta$.
\end{lemma}
\begin{proof}
There are $N=\sum_{i=0}^{n-1} 2^i = 2^n -1$ binary strings
of length less than $n$. A fortiori there are at most $N$
elements of $S$ that can be computed by
binary programs of length less than $n$, given $y$.
This implies that at least $m-N$  elements of $S$ cannot
be computed by binary programs of length less than $n$, given $y$.
Substituting $n$ by  $\lfloor \log m \rfloor  - \delta$ together
with Definition~\ref{def.KC} yields the lemma.
\end{proof}

If we are given $S$ as an explicit table
then we can simply enumerate its elements
(in, say, lexicographical order) using a fixed program not depending
on $S$ or $y$. Such a fixed program can be given in $O(1)$ bits.
Hence we can upper bound the complexity as $C(x|S,y) \lea \log |S| $.

{\bf Incompressibility Method:}
One reason to formulate a notion of quantum Kolmogorov complexity,
apart from its interpretation as the information in an individual
quantum state, is the following.
We hope to dupplicate the success
of the classical version as
a proof method, the incompressibility method, in the theory of
computation and combinatorics \cite{LiVi97}:
In a typical proof using the incompressibility method,
one first chooses an incompressible object from the
class under discussion.
The argument invariably says that if a desired property
does not hold, then in contrast with the assumption, the object
can be compressed. This yields the required contradiction.
Since most objects are almost incompressible, the desired property
usually also holds for almost all objects, and hence on average.
The hope is that one can use the quantum Kolmogorov
complexity to show, for example, lower
bounds on the complexity of quantum computations.

{\bf Prefix Kolmogorov complexity:}
For technical reasons we also need a variant of complexity,
so-called prefix complexity, which associated with Turing machines
for which the set of programs resulting in a halting computation
is prefix-free. We can realize this by equipping the Turing
machine with a read-only input tape which is read from left-to-right
without backing up, a separate read/write work tape,
an auxiliary read-only input tape,
and a write-only output tape that is written from left-to-right
without backing up. All tapes are one-way infinite. Such Turing
machines are called {\em prefix machines}
since the set of halting programs for such a machine
forms a prefix-free set.
Taking the universal prefix machine $U$ we can define
the prefix complexity analogously with the plain Kolmogorov complexity.
Let $x^*$ be the shortest program for $x$ that is enumerated first
in a fixed general enumeration process (say, by dovetailing
the running of all candidate programs)  of all programs for which the
reference universal prefix machine computes $x$.  Then, the set
$\{x^* : U(x^*)=x, x \in \{0,1\}^*\}$ is a {\em prefix code}.
That is, if $x^*$
and $y^*$ are code words for $x$ and $y$, respectively,
 with $x \neq y$, then $x^*$ is not
a prefix of $y^*$.

Let $\langle \cdot \rangle$ be a standard invertible
effective one-one encoding from ${\cal N} \times {\cal N}$
to prefix-free recursive subset of ${\cal N}$.
For example, we can set $\langle x,y \rangle = x'y'$.
We insist on prefix-freeness and
recursiveness because we want a universal Turing
machine to be able to read an image under $\langle \cdot \rangle$
from left to right and
determine where it ends.
Let $P_1 ,P_2 , \ldots$ be a standard enumeration
of all prefix machines, and let $\phi_1 , \phi_2 , \ldots$
be the enumeration of corresponding functions
that are computed:
$P_i$ computes $\phi_i$.
It is easy to see that (up to the prefix-free
encoding) these functions are exactly
the {\em partial recursive} functions
or {\em computable} functions.
The {\em conditional complexity} of $x$ given $y$
with respect to a prefix machine $P$ is
\[K_P(x|y) = \min_{p \in \{0,1\}^*} \{l(p): P (\langle p,y\rangle)=x \}. \]
The unconditional complexity of $x$ with respect to $P$ is defined
by $K(x) = K(x| \epsilon )$.
Choose a universal prefix machine
$UP$ that expresses its universality in the following manner:
\[UP(\langle \langle i,p \rangle ,y \rangle ) =
P_i (\langle p,y\rangle) \]
 for all $i$ and $ p,y$.
Proving the Invariance Theorem for prefix machines goes
by the same reasoning as before. Then, we can define:

\begin{definition}\label{def.KolmK}
Fix a $UP$ as above
 as our {\em reference universal prefix computer}, and define
the {\em conditional prefix complexity} of $x$ given $y$
by
\[K(x|y) = \min_{p \in \{0,1\}^*} \{l(p): UP (\langle p,y\rangle)=x \}. \]
The unconditional Kolmogorov complexity of $x$ is defined
by $K(x) = K(x| \epsilon )$.
\end{definition}

Note that $K(x|y)$ can be slightly larger than $C(x|y)$, but for
all $x,y$ we have
\[ C(x|y) \lea K(x|y) \lea C(x|y)+ 2 \log C(x|y) . \]
For example, the incompressibility laws hold also for $K(x)$
but in slightly different form.
The nice thing about $K(x)$ is that we can interpret $2^{-K(x)}$
as a probability distribution since $K(x)$ is the length of
a shortest prefix-free program for $x$. By the fundamental
Kraft's inequality, see for example \cite{CT91,LiVi97}, we know that
if $l_1 , l_2 , \ldots$ are the code-word lengths of a  prefix code,
then $\sum_x 2^{-l_x} \leq 1$. This leads to the notion
of the ``universal distribution'' ${\bf m}(x) = 2^{-K(x)}$
that assigns high probability to simple objects (that is, with
low prefix complexity) and low probability to complex objects
(that is, with high prefix complexity)---a rigorous form of Occam's Razor.

\section{Appendix: Quantum Turing Machines}
\label{app.B}

We base quantum Kolmogorov complexity on quantum Turing
machines.
The simplest way to explain the idea quantum computation
is perhaps by way of
probabilistic (randomized) computation. This we explain
here. Then, the definition of the quantum (prefix) Turing machine
is given in the main text in Section{sect.model}.

\subsection{Notation}
For every $N$ the finite-dimensional Hilbert space ${\cal H}_N$ 
has a canonical basis $\ket{e_0}, \ldots , \ket{e_{N-1}}$. Assume
that the canonical basis of ${\cal H}_N$ is also the beginning
of the canonical basis of ${\cal H}_{N+1}$. The $m$-fold tensor
product $\otimes_{i=1}^m {\cal H}$ of a Hilbert space ${\cal H}$ is denoted by
${\cal H}^{\otimes m}$. 

%Let $\{0,1\}^n$ be the set of binary strings
%of length $n$. If $x \in \{0,1\}^n$ then $x=x_1 \ldots x_n$
%and we write $l(x)=n$. The $n$-fold tensor product 
%${\cal Q}^{\otimes n}$  is the Hilbert space of $n$ qubits. We have
%$\ket{x}= \otimes_{i=1}^n \ket{x_i}$ is an element of ${\cal Q}^{\otimes n}$
%and we identify ${\cal Q}^{\otimes n}$ with ${\cal H}_N$ ($N=2^n$)
%and the canonical basis element $\ket{e_x} = \ket{x}$. 

A pure quantum state 
$\phi$ represented as a unit length vector in such a Hilbert space
is denoted as $\ket{\phi}$ and the corresponding element of the dual space
(the conjugate transpose) is written as 
$\phi^{\dagger}$ or $\bra{\phi}$. The inner product
 of $\bra{\phi}$ and $\ket{\psi}$ is written in physics notation
as $\bracket{\phi}{\psi}$ and in mathematics notation 
as $\phi^{\dagger} \psi$.
The ``bra-ket'' notation
is due to P. Dirac and is the standard
quantum mechanics notation.
The ``bra''  $\bra{x}$ denotes a row vector with complex entries, and
``ket'' $\ket{x}$ is the column vector consisting of the
conjugate transpose of  $\bra{x}$ (columns
interchanged with rows and the imaginary part of the entries
negated, that is, $\sqrt{-1}$ is replaced by $- \sqrt{-1}$).

Of special importance is the two-dimensional
Hilbert space 
${\cal C}^2$, where ${\cal C}$
is the set of complex real numbers, and $\ket{0}$, $\ket{1}$ is its canonical
orthonormal basis. An element of ${\cal C}^2$ is called a {\em qubit}
(quantum bit in analogy with an element of $\{0,1\}$ which is called
a {\em bit} for ``binary digit'').  
To generalize this to strings of $n$ qubits,
we consider the quantum state space
${\cal C}^N$ with $N=2^n$. The basis vectors $e_0 , \ldots , e_{N-1}$
of this space are parametrized by binary strings of length $n$, so that
$e_0$ is shorthand for $e_{0 \ldots 0}$ and $e_{N-1}$ is shorthand
for $e_{1 \ldots 1}$.
Mathematically, ${\cal C}^N$ is decomposed into a tensor product
of $n$ copies of ${\cal C}^2$, written as $({\cal C}^2)^{\otimes n}$,
 and an $n$-qubit state
$\ket{a_1 \ldots a_n}$ in bra-ket notation can also be written
as the tensor product $\ket{a_1} \otimes \ldots \otimes \ket{a_n}$,
or shorthand as $\ket{a_1} \ldots \ket{a_n}$, a string
of $n$ qubits, the qubits being distinguished
by position.

\subsection{Probabilistic Computation} 
Consider the well known
probabilistic Turing machine which is just like an ordinary
Turing machine, except that at each step the machine can make
a probabilistic move which consists in flipping a (say fair)
coin and depending on the outcome changing its state to either
one of two alternatives. This means that at each such
probabilistic move the computation of the machine splits into
two distinct further computations each with probability $\frac{1}{2}$.
Ignoring the deterministic computation steps,
a computation involving $m$ coinflips can be viewed as
a  binary computation tree of depth $m$ with $2^m$ leaves,
where the set of nodes at level $t \leq m$ correspond to the possible states
of the system after $t$ coinflips, every state occurring with
probability $1/2^t$. For convenience, we can label
the edges connecting a state $x$ directly with a state $y$
with the probability that a state $x$ changes into state $y$
in a single coin flip (in this example all
edges are labeled `$\frac{1}{2}$').

For instance, given an arbitrary Boolean formula containing
$m$ variables, a probabilistic machine can flip its coin $m$
times to generate each of the $2^m$ possible truth assignments
at the $m$-level nodes,
and subsequently check in each node deterministically wether the 
local assignment
makes the formula true. If there are $k$ distinct such assignments then
the probabilistic machine finds that the formula
is satisfiable with probability at least $k/2^m$---since there
are $k$ distinct computation paths leading to a satisfiable
assignment.

Now suppose the probabilistic machine is
hidden in a black box and the computation
proceeds without us knowing
the outcomes of the coin flips.
Suppose that after $m$ coin flips
we open part
of the black box and observe the bit
which denotes the truth assignment
for variable $x_5$ ($5 \leq m$).
Before
we opened the black box all $2^m$ initial truth assignments to variables
$x_1, \ldots , x_m$ were still equally possible, each with probability $1/2^m$.
After we observed the state of variable $x_5$, say 0, the probability
space of possibilities has collapsed to the truth assignments
which consist of all binary vectors with a 0 in the 5th position
each of which has probability renormalized to $1/2^{m-1}$.

\subsection{Quantum Computation}
\label{sec.cflip}
A quantum Turing machine can be viewed as a generalization
of the probabilistic
Turing machine. Consider the same computation tree.
In the probabilistic computation
there is a probability  $p_i \geq 0$ associated with each node $i$ (state
of the system) at the same level in the tree,
such that $\sum p_i = 1$, summed over the nodes at the same level.
In a quantum mechanical computation
there is a ``probability amplitude'' $\alpha_i$
associated with each basis state $\ket{i}$
of the system.
Ignore for the moment the quantum equivalent
of the probabilistic coin flip to produce 
the computation tree. Consider the simple case
(corresponding to the probabilistic example of the states of
the nodes at the $m$th level of the computation tree) where $i$ runs through
the classical values  0
through $2^m-1$, in the quantum case
represented by the orthonormal basis $m$-qubit states 
$\ket{00 \ldots 0}$ through $\ket{11 \ldots 1}$.
Then, the nodes at level $m$ are in a superposition 
$\ket{\psi} = \sum_{i  \in \{0,1\}^m} \alpha_i \ket{i}$
with the probability amplitudes satisfying 
$\sum_{i \in \{0,1\}^m} \norm{\alpha_i}^2 =1$. 
 
The amplitudes are complex numbers
satisfying $\sum \norm{\alpha_i}^2 = 1$, where if $\alpha_i = a + b \sqrt{-1}$
then $\norm{\alpha_i} = \sqrt{a^2 + b^2}$, and the summation is taken over all
distinct states of the observable at a particular instant. 
We say ``distinct'' states since the quantum mechanical calculus
dictates that equal states are grouped together: If state $\ket{\phi}$
of probability amplitude $\alpha$ equals state $\ket{\psi}$
of probability amplitude $\beta$, then their combined contribution in the
sum is $\norm{\alpha + \beta}^2 \ket{\phi}$.
The transitions are governed by a matrix $U$
which represents the program being executed. Such a program has to satisfy
the following constraints. Denote the set of possible configurations
of the Turing machine by $X$, where
$X$ is the set of $m$-bits column vectors (the basis states) for simplicity.
Then $U$ maps
the column vector $\underline{\alpha} = (\alpha_x)_{x \in X}$
to $U \underline{\alpha}$. Here $\underline{\alpha}$ is a $(2^m)$-element
complex vector
of amplitudes of the quantum superposition of
the $2^m$ basis states before the step, and $U \underline{\alpha}$
the same after the step concerned. The special property
which $U$ needs to satisfy in quantum mechanics is that it is
{\em unitary}, that is, $U^{\dagger}  U = I$ where $I$
is the identity matrix and $U^{\dagger}$ is the conjugate transpose
of $U$ (as with the bra-ket, ``conjugate''
means that all $\sqrt{-1}$'s are replaced by $- \sqrt{-1}$'s
and `transpose' means that the rows and columns are interchanged).
In other words, $U$ is unitary iff $U^{\dagger} = U^{-1}$.

The unitary constraint on the evolution of the computation
enforces two facts.
\begin{enumerate}
\item
If $U^0 \underline{\alpha} = \underline{\alpha}$ and $U^t = U  U^{t-1}$
then
$\sum_{x \in X} \norm{(U^t \underline{\alpha})_x}^2 = 1$  for all $t$ 
(discretizing
time for convenience).
\item
A quantum computation is reversible (replace $U$ by $U^{\dagger} = U^{-1}$).
In particular this means that a computation 
$U^t  \underline{\alpha_0} =  \underline{\alpha_t}$ is undone by
running the computation stepwise in reverse:
${U^{\dagger}}^t  \underline{\alpha_t} =  \underline{\alpha_0}$.
\end{enumerate}

The quantum version of a single bit
is a superposition of the two basis states a classical bit:
\[ \ket{\psi} = \alpha \: \zero + \beta  \:\one, \]
 where
\mbox{$\norm{\alpha}^2 +\norm{\beta}^2 =1$}. Such a state
$\ket{\psi}$ is called a quantum bit or {\em qubit}. It consists
of partially the basis state $\ket{0}$
and partially the basis state $\ket{1}$.
The states are denoted by the column vectors
of the appropriate complex probability amplitudes. For the
basis states the vector notations are:
$\ket{0} = { 1 \choose 0}$ (that is, $\alpha = 1$ and
$\beta = 0$), and $\ket{1} = {0 \choose 1}$ (that is, $\alpha = 0$
and $\beta = 1$). We also write $\ket{\phi}$ as the
column vector $\underline{\phi} = {{\alpha} \choose {\beta}}$.

Physically, for example, the state $\ket{\psi}$ can be the state
of a polarized photon, and the basis states are horizontal or vertical
polarization, respectively.
Upon measuring according to the basis states, that is, passing
the photon through a
medium that is polarized either in the horizontal or vertical
orientation, the photon is observed
with probability $\norm{\alpha}^2$ or probability
$\norm{\beta}^2 $, respectively.
Consider a sample computation on a one-bit
computer executing the
unitary operator:
\begin{equation}\label{eq.Sqrn}
S ~=~ \frac{1}{\sqrt{2}} \,
    \left( \begin{array}{rr} 1&1\\-1&1 \end{array} \right) .
\end{equation}
It is easy to verify, using common matrix calculation, that
\begin{eqnarray*}
 S \ket{0} & = & \frac{1}{\sqrt{2}}  \:\ket{0} - \frac{1}{\sqrt{2}}  \:\ket{1} ,
\;
 S \ket{1}  =  \frac{1}{\sqrt{2}}  \:\ket{0} + \frac{1}{\sqrt{2}}  \:\ket{1}, \\
 S^2 \ket{0} & = & 0  \:\ket{0} - 1  \:\ket{1} = - \ket{1}, \;
 S^2 \ket{1}  =  1  \:\ket{0} + 0  \:\ket{1} =  \ket{0} . \\
\end{eqnarray*}
If we observe
the computer in state $S \ket{0}$, then
the probability of observing state $\ket{0}$
is $(\frac{1}{\sqrt{2}})^2 = \frac{1}{2}$,
and the probability to observe $\ket{1}$ is $(- \frac{1}{\sqrt{2}})^2 = \frac{1}
{2}$. However,
if we observe
the computer in state $S^2 \ket{0}$, then
the probability of observing state $\ket{0}$
is $0$,
and the probability to observe $\ket{1}$ is $1$. Similarly,
if we observe
the computer in state $S \ket{1}$, then
the probability of observing state $\ket{0}$
is $(\frac{1}{\sqrt{2}})^2 = \frac{1}{2}$,
and the probability to observe $\ket{1}$ is
$(  \frac{1}{\sqrt{2}})^2 = \frac{1}{2}$.
If we observe
the computer in state $S^2 \ket{1}$, then
the probability of observing state $\ket{0}$
is $1$,
and the probability to observe $\ket{1}$ is $0$. Therefore, the operator $S$
inverts a bit when it is applied twice in a row, and hence
has acquired the charming name {\em square root of `not'}.
%It is a simple exercise to write $S$ in terms of an
%if--then--else program:
%\begin{prog}
%\>{\bf if} $\ket{\Psi}=\zero$ \={\bf then} \=$\ket{\Psi} =
 %\frac{1}{\sqrt{2}} \: \ket{0} - \frac{1}{\sqrt{2}} \: \ket{1}$ \\[0.5ex]
%%\left\{ \begin{array}{ll} \zero & \mbox{with amplitude $+\frac{1}{\sqrt{2}}$}
%\\[0.5ex]
                          %%\one & \mbox{with amplitude $-\frac{1}{\sqrt{2}}$}
        %%\end{array} \right.$\\*[1.5ex]
%\>\>{\bf else} \>$\ket{\Psi} =
 %\frac{1}{\sqrt{2}} \: \ket{0} + \frac{1}{\sqrt{2}} \: \ket{1}$  \\[0.5ex]
%%\left\{ \begin{array}{ll} \zero & \mbox{with amplitude $+\frac{1}{\sqrt{2}}$}
%\\[0.5ex]
                          %%\one & \mbox{with amplitude $+\frac{1}{\sqrt{2}}$}
        %%\end{array} \right.$
%\end{prog}
%Note that there is no classical unconditional
%Boolean operator that has the effect of $S$;
%this is one difference
%between quantum computation and probabilistic computation.
%
%In computing the above amplitudes, subsequent to two applications of
%$S$, according to matrix calculus we found that
%\begin{eqnarray*}
%S^2 \ket{1} &=& \textstyle \frac{1}{\sqrt{2}} \left(
%\frac{1}{\sqrt{2}}(\zero-\one)+\frac{1}{\sqrt{2}}(\zero+\one) \right)\\*[0.5ex]
%&=& \textstyle \frac{1}{2} (\zero-\one+\zero+\one) ~=~ \zero.
%\end{eqnarray*}
In contrast, with the analogous probabilistic calculation, 
flipping a coin two times in a row,
we would have found that the probability of each computation path
in the complete binary computation tree of depth 2 was $\frac{1}{4}$, and
the states at the four leaves of the tree were $\ket{0}, \ket{1},
\ket{0}, \ket{1}$, resulting in a total probability of
observing $\ket{0}$ being $\frac{1}{2}$, and the total probability of observing
$\ket{1}$ being $\frac{1}{2}$ as well.

The quantum principle involved in the above example
is called {\em interference}, similar to the related
light phenomenon in the seminal ``two slit experiment:''
If we put a screen  with a single small enough hole in
between a light source and a target,
then we observe a gradually dimming illumination of the target screen,
the brightest spot being colinear with the light source and the hole.
If
we put a screen with {\em two} small holes in between, then
we observe a diffraction pattern of bright and dark stripes
due to interference. Namely, the light hits every point on the screen
via two different routes (through the two different holes).
If the two routes differ by an even number of half wave lengths,
then the wave amplitudes at the target are added, resulting in
twice the amplitude and a bright
spot, and if they differ by an odd number of half wave lengths
then the wave amplitudes are in opposite phase and are subtracted
resulting in zero and a dark spot. Similarly, with quantum
computation, if the quantum state is
$ \ket{\psi} = \alpha \ket{x} + \beta \ket{y} $,
then for $x=y$ we have a probability of observing
$\ket{x}$ of $\norm{\alpha + \beta}^2$, rather than
$\norm{\alpha}^2 + \norm{\beta}^2$ which it would have been
in a probabilistic fashion. For example, if $\alpha = \frac{1}{\sqrt{2}}$
and $\beta = - \frac{1}{\sqrt{2}}$ then the probability of observing
$\ket{x}$ is 0 rather than $\frac{1}{2}$, and with the sign of $\beta$
inverted we observe $\ket{x}$ with probability 1.

\subsection{Quantum Algorithmics}
A quantum algorithm corresponds to a unitary transformation $U$
that is built up from elementary unitary transformations,
every one of which only acts on one or two qubits.
The algorithm applies $U$ to an initial classical state containing
the input and then makes a final measurement to extract the output from
the final quantum state.
The algorithm is ``efficient'' if the number of elementary operations is
``small'', which usually means at most polynomial in the length of the input.
Quantum computers can do everything a classical computer
can do probabilistically ---
and more.

We are now in the position to
explain 
the quantum equivalent of
a probabilistic coin flip as promised in Section~\ref{sec.cflip}. 
This is a main trick enhancing the power of quantum computation.
A sequence of $n$ fair coin flips
``corresponds'' to a sequence $H_n$ of
$n$ one-qubit unitary operations, the Hadamard transform,
\[ H ~=~ \frac{1}{\sqrt{2}} \,
    \left( \begin{array}{rr} 1&1\\1&-1 \end{array} \right)
\]
on the successive bits of a register of $n$ bits
originally in the all--0 state $\ket{\psi}= \ket{00 \ldots 0}$.
The result is a superposition
\[ H_n \ket{\psi} = \sum_{x \in \{0,1\}^n} 2^{-n/2} \:\: \ket{x}\]
of all the $2^n$
possible states of the register, each with amplitude $2^{-n/2}$
(and hence probability of being observed of $2^{-n}$).

The Hadamard transform is ubiquitous in quantum computing;
its singlefold action is similar to that of the transform $S$
of (\ref{eq.Sqrn}) with the  the roles of ``0'' and ``1''
partly interchanged. In contrast to
$S^2$ that implements the logical ``not,''
we have $H^2 = I$ with $I$ the identity matrix.

Subsequent to application of $H_n$,
the computation proceeds in parallel along the
exponentially many computation paths in quantum coherent superposition.
A sequence of tricky further unitary operations, for example
the ``quantum Fourier transform,'' and observations
serves to exploit interference (and so-called entanglement) phenomena
to effect a high probability of eventually observing outcomes
that allow us to determine the desired
result, and suppressing the undesired spurious outcomes.

One principle that is used in many quantum
algorithms is as follows.
If $A$ is a classical algorithm for computing some function $f$,
possibly even irreversible like $f(x)  \equiv  x \pmod 2$,
then we can turn it into a  unitary transformation which maps classical
state $\ket{x,0}$ to $\ket{x,f(x)}$.
Note that we can apply $A$ to a superposition of all $2^n$ inputs:
$$
A\left(2^{-n/2}\sum_x\ket{x,0}\right)=
2^{-n/2}\sum_x\ket{x,f(x)}.
$$
In some sense this state contains the results of computing $f$
for {\em all} possible inputs $x$, but we have only
applied $A$ once to obtain it.
This effect together with the interference phenomenon
is responsible for one of the advantages of quantum over classical
randomized computing and is called {\em quantum parallelism}.

This leaves the question of how the input to a computation
is provided and how the output is obtained.
Generally, we restrict ourselves to the case where the quantum computer has
a classical input. If the input $x$ has $k$ bits, and the number
of qubits used by the computation is $n \geq k$ (input plus work space),
then we
pad the input with nonsignificant 0's and start the quantum
computation in an initial state (which must be in ${\cal C}^N$)
$\ket{x0 \ldots 0}$. When the computation finishes the resulting
state is a unit vector in ${\cal C}^N$, say $\sum_i \alpha_i \ket{i}$
where $i$ runs through $\{0,1\}^n$ and the probability
amplitudes $\alpha_i$'s satisfy $\sum_i \norm{\alpha_i}^2 =1$.
The output is obtained by performing a measurement with as possible outcomes
the basis vectors. The observed output is probabilistic: we observe
basis vector $\ket{i}$, that is, the $n$-bit string $i$, with probability
$\norm{\alpha_i}^2$. 

\section*{Acknowledgement}
The ideas presented in this paper were developed from 1995 through early 1998.
Other interests prevented me from earlier publication.
I thank Harry Buhrman, Richard Cleve, Wim van Dam, P\'eter G\'acs,
 Barbara Terhal, John Tromp,
Ronald de Wolf, and a referee for
discussions and comments on QKC.
 
\bibliographystyle{IEEE}

\newpage
\begin{biography}{Paul M.B. Vit\'anyi}
received his Ph.D. from the Free University 
of Amsterdam (1978).  He holds positions at the national 
CWI research institute in Amsterdam, and he is professor of computer science
at the University of Amsterdam.  He serves on the editorial boards
of Distributed Computing, Information Processing Letters, 
Theory of Computing Systems, Parallel Processing Letters,
Journal of Computer and Systems Sciences (guest editor),
and elsewhere. He has worked on cellular automata, 
computational complexity, distributed and parallel computing, 
machine learning and prediction, physics of computation, 
and Kolmogorov complexity. Together with Ming Li 
they pioneered applications of Kolmogorov complexity 
and co-authored ``An Introduction to Kolmogorov Complexity 
and its Applications,'' Springer-Verlag, New York, 1993 (2nd Edition 1997),
parts of which have been translated into Chinese,
Russian and Japanese.

\end{biography}

\end{document}